\documentclass[twocolumn,twocolappendix]{aastex631}

\usepackage{amsmath,hyperref}
\usepackage{color}

\def\bi#1{\mbox{\boldmath{$#1$}}}
\defcitealias{Soares_2022}{S22}

\makeatletter
\DeclareRobustCommand{\HI}{%
  \mbox{H\:\check@mathfonts\fontsize\sf@size\z@\selectfont I}%
}
\makeatother

\received{XXX}
\revised{YYY}
\accepted{ZZZ}
\submitjournal{ApJ}

\shorttitle{Spatial Variations in 21 cm GP}

\shortauthors{Diao, Grumitt \& Mao}

\begin{document}

\title{Modeling Foreground Spatial Variations in 21~cm Gaussian Process Component Separation}

\author[0000-0001-7301-2318]{Kangning Diao}
\affiliation{Department of Astronomy, Tsinghua University, Beijing 100084, China}

\author[0000-0001-9578-6111]{Richard D.P. Grumitt}
\affiliation{Department of Astronomy, Tsinghua University, Beijing 100084, China}

\author[0000-0002-1301-3893]{Yi Mao}
\affiliation{Department of Astronomy, Tsinghua University, Beijing 100084, China}

\correspondingauthor{Richard D.P. Grumitt, Yi Mao}
\email{rgrumitt@tsinghua.edu.cn (RDPG), ymao@tsinghua.edu.cn (YM)}



\begin{abstract}

{The 21~cm line from neutral hydrogen captures rich information in the Universe. However, the 21~cm signals are highly contaminated by foreground emission, making accurate component separation vital for cosmological analyses.} Gaussian processes (GPs) have been utilized for {component separation} in 21~cm data analyses, {exploiting} the distinct spectral behavior of the cosmological and foreground signals, which are modeled through the GP covariance kernel. Previous approaches have employed a global GP kernel along all lines of sight (LoS). In this work, we study Bayesian approaches that allow for spatial variations in foreground kernel parameters, testing them against simulated \HI{} intensity mapping observations, {which measures the total \HI{} intensity in large voxels}. We consider a no-pooling (NP) model, treating each LoS independently by fitting for separate covariance kernels, and a hierarchical Gaussian Process (HGP) model, allowing for variation in kernel parameters between different LoS, regularized through a global hyperprior. We find that {including} spatial variations in the GP kernel parameters results in a significant improvement in cosmological signal recovery, achieving up to a 30\% reduction in the standard deviation of the residual distribution. Allowing for spatial variations also improves the recovery of the \HI{} power spectra and wavelet scattering transform coefficients. Whilst the NP model achieves superior recovery as measured by the residual distribution, it {is computationally demanding}, faces convergence challenges, and is prone to overfitting. The HGP model strikes a balance between the accuracy and robustness. Further improvements to the HGP model will require more physically-motivated modeling of foreground spatial variations.  Our code can be found in \href{https://github.com/dkn16/H21F}{this repository}.

\end{abstract}

\keywords{Cosmology(343) --- Large-scale structure of the universe(902) --- 
21 cm lines(690) --- Gaussian Processes regression(1930) }


\section{Introduction}

Measurements of the 21~cm signal from neutral hydrogen provide a powerful cosmological probe. At high redshifts ($z\gtrsim 6$), these observations are a key window into the astrophysics of the Epoch of Reionization (EoR) and Cosmic Dawn (CD), as well as allowing for independent constraints on cosmological parameters \citep{2012RPPh...75h6901P, Furlanetto2016}. At lower redshifts, neutral hydrogen (\HI{}) intensity mapping (IM) can be used as a tracer for the large-scale structure of the Universe \citep{2001JApA...22...21B, 2004MNRAS.355.1339B, 2008PhRvL.100i1303C}. 

In this work, we focus on \HI{} IM observations. \HI{} IM aggregates unresolved 21 cm emissions from low-aperture radio telescopes, obviating the need for high-resolution imaging of \HI{} emission from individual galaxies. In the post-reionization Universe, \HI{} is largely confined within dark matter halos \citep{2018ApJ...866..135V}, {meaning} the 21 cm radiation spontaneously emitted by \HI{} gas {traces the cosmological density field}. Many radio telescopes targeting \HI{} IM are planned {or under construction, such as the Packed Ultra-wideband Mapping Array \citep[PUMA,][]{2019BAAS...51g..53S}, Canadian Hydrogen Observatory and Radio-transient Detector \citep[CHORD,][]{2019clrp.2020...28V}, and Square Kilometre Array \citep[SKA,][]{2020PASA...37....2W},} or have been built, {such as the Canadian Hydrogen Intensity Mapping Experiment \citep[CHIME,][]{2022ApJS..261...29C}, Tianlai experiment \citep{2012IJMPS..12..256C}, and MeerKAT telescope \citep[e.g.][]{2016mks..confE..32S}. Among these telescopes, the MeerKAT telescope, which consists of 64 dishes, has already delivered high quality observations both with \textit{interferometry} \citep{2023arXiv230111943P}, where the high resolution allows us to probe small-scale fluctuations, and \textit{single-dish} mode \citep[e.g.][]{2023MNRAS.518.6262C}, which is able to access large cosmological scales and has simple noise properties. A key science project of MeerKAT is the MeerKAT Large Area Synoptic Survey \citep[MeerKLASS, e.g.][]{2024arXiv240721626M}, which uses 64 dishes in 41 repeated single-dish scans to measure baryon acoustic oscillations via the 21 cm IM technique, aiming to constrain the nature of dark energy.}

However, the accurate detection and characterization of such cosmological signals face a severe challenge in foreground removal. Any 21~cm signal will be orders of magnitude weaker than the foreground emission, consisting of extragalactic emission along with diffuse and partially polarized Galactic emission \citep{2002ApJ...564..576D, 2003MNRAS.346..871O, 2006ApJ...650..529W, 2008MNRAS.389.1319J, 2009A&A...500..965B, 2010A&A...522A..67B, 2011PhRvD..83j3006L, 2012MNRAS.419.3491L, 2020MNRAS.495.2813G, 2020PASP..132f2001L}. 

{Foreground removal methods typically proceed by assuming that the foreground emission varies much more {smoothly} with frequency and has a much greater amplitude than the 21 cm signal \citep{2011PhRvD..83j3006L, PhysRevD.91.083514, 2018MNRAS.478.3640M}. For \HI{} IM the most widely used methods are blind component separation (BCS) algorithms, e.g.\ principal component separation (PCA) \citep{2013ApJ...763L..20M, 2015MNRAS.447..400A,2015MNRAS.454.3240B, Cunnington2021} that is equivalent to the singular value decomposition \citep{2015ApJ...815...51S}, and its variants including robust PCA \citep{2019AJ....157....4Z}, singular vector projection (a semi-blind PCA) \citep{2023ApJ...945...38Z}, independent component analysis \citep{2012MNRAS.423.2518C,2014MNRAS.441.3271W,2015MNRAS.447..400A} and generalized morphological
component analysis \citep{2013MNRAS.429..165C, 2020MNRAS.499..304C}. Broadly, these methods assume that the observed sky signal may be written as $\bi{X}=\bi{A}\bi{S}+\bi{N}$, where $\bi{X}$ is an $N\times p$ matrix representing the observed data for the $N$ frequency channels and $p$ pixels in each map, $\bi{S}$ is an $N_{\mathrm{fg}}\times p$ foreground signal matrix for the $N_\mathrm{fg}$ foreground components, $\bi{A}$ is an $N\times N_\mathrm{fg}$ mixing matrix, and $\bi{N}$ is an $N\times p$ matrix that includes both cosmic signal and noise. Whilst each method imposes different constraints on the foreground signal, they do not impose any strict parametric form on the foreground {spectral energy distribution (SED)}, and each seeks to solve an optimization problem to determine values for $\bi{A}$ and $\bi{S}$, such that the foreground cleaned signal can be obtained from the residual, $\bi{\epsilon}=\bi{X}-\bi{A}\bi{S}$.}

{Beyond these BCS approaches that optimize for $\bi{A}$ and $\bi{S}$, Gaussian processes (GPs) have seen extensive applications for non-parametric 21 cm component separation analyses. This includes applications to EoR simulations \citep{2018MNRAS.478.3640M, 2024MNRAS.527.3517M, 2019MNRAS.488.4271G, 2019MNRAS.484.2866O,  2021MNRAS.501.1463K, 2023MNRAS.524.3724C} and to real data \citep{2021MNRAS.500.2264H, 2020MNRAS.495.2813G, 2020MNRAS.493.1662M}. In \citealt{Soares_2022} (hereafter \citetalias{Soares_2022}), it was shown that GPs outperform PCA on small scales when recovering the \HI{} power spectrum, in the context of single dish \HI{} IM simulations. GPs provide a non-parametric model for the line of sight (LoS) behavior of the foreground and \HI{} components}, belonging to a class of non-parametric Bayesian models, where one assigns a prior over the space of functions that describe the emission component SEDs \citep{RasmussenW06}. The distinct spectral behaviors of the emission components can be accounted for through the GP covariance kernel, enabling a flexible component separation. One can also fully marginalize the GP model parameters, allowing for a full propagation of uncertainty through the component separation process \citep{flaxman2015fast, lalchand2020approximate}.

In this work, we consider adapting the GP component separation framework to allow spatial variations in the foreground kernel parameters, so that we can better model spatial variations in the foreground spectral properties. {This has previously been found to be a significant issue for cosmic microwave background (CMB) component separation, where foregrounds at the $\mathrm{GHz}$ frequency have been shown to display significant spatial variation} \citep{2016A&A...594A..10P, 2018JCAP...04..023R, 2020MNRAS.495..578J}. In previous work, the foreground GP kernel parameters have been assumed to be identical along every LoS, i.e. a ``complete pooling'' model (e.g. \citetalias{Soares_2022}). We study the impact of allowing spatial variations through a ``no pooling'' model, where the foreground kernel parameters are allowed to vary independently along \emph{every} LoS, and a hierarchical model where the kernel parameters are allowed to vary but are assumed to be drawn from some underlying hyper-distribution, whose parameters are jointly fit during the inference process. The hyper-distribution acts to regularize pixel-to-pixel variations in GP kernel parameters and allows for the sharing of information between different LoS \citep{doi:10.1198/004017005000000661, gelman2006data}. Hierarchical foreground modeling has previously been used in the context of parametric CMB component separation in \cite{2020MNRAS.496.4383G}.

The structure of the rest of this paper is as follows. In Section \ref{sec: gp models} we give an overview of GP models, their specific application to 21~cm component separation, and the form of the various GP models we consider in this work. In Section \ref{sec: sims} we describe the simulated \HI{} IM observations,  against which we test our component separation method, consisting of a mock MeerKLASS-type survey \citep{2016mks..confE..32S}, and the computational setup used to perform inference over our models. In Section \ref{sec:results} we describe the results from these simulations, including the uncompressed pixel-level recovery and predictive performance, along with the recovery of \HI{} field summary statistics. In Section \ref{sec:dis} we discuss notable modeling and recovery details, covering the effect of bias correction on power spectrum recovery, the modeling choices used to define the hierarchical GP models, and the GP inference approach used in this work. In Section \ref{sec: conclusions} we summarize our method and results and discuss improvements that can be made in the modeling of foreground spatial variations for component separation. {We provide additional technical details in the appendices, with Appendix \ref{subsec: gp overview} giving a general overview of GP modeling and inference, Appendix \ref{subsec:vis_insp} showing visual results demonstrating the signal recovery of various GP models over a range of frequencies, and Appendix \ref{sec: cp2 performance} providing additional discussion regarding the poor performance of the CP2 model (defined in Section \ref{subsec: gp sep setup}).}

\section{Gaussian Processes}\label{sec: gp models}

\begin{figure*}
    \centering
    \includegraphics[width=\linewidth]{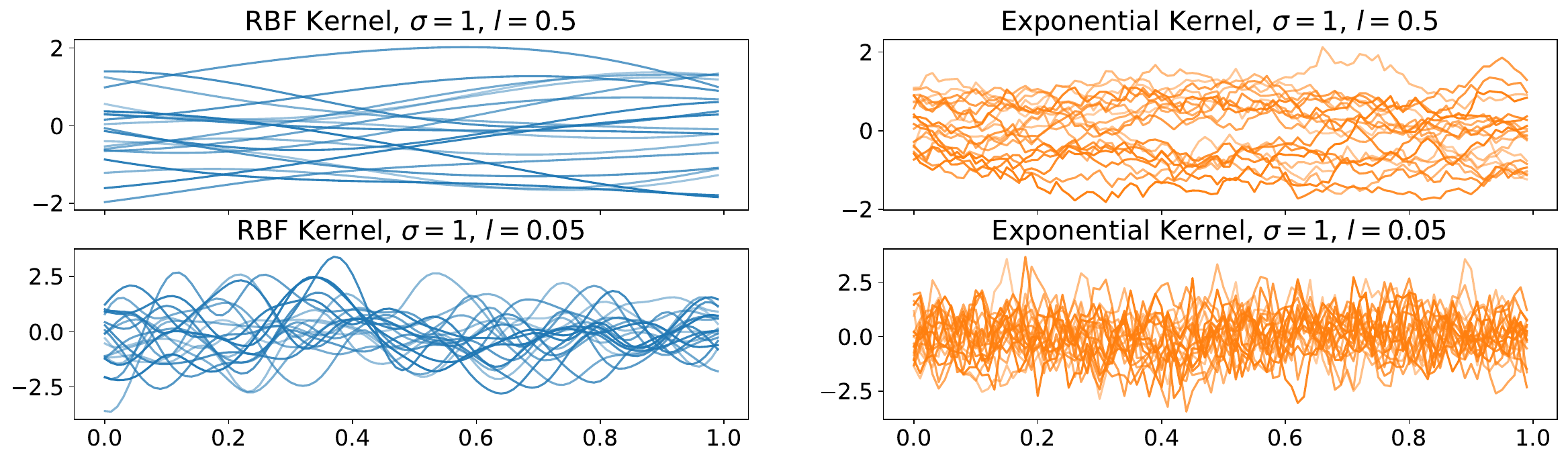}
    \caption{Samples from different GP kernels with different length scales $l$. {The $x$-axis represents the one-dimensional function domain ($x\in[0, 1]$), with the $y$-axis giving the values of the GP function realizations over this domain.} It can be seen that decreasing the length scale results in more rapidly varying GP realizations. Functions allowed by the RBF kernel are infinitely differentiable, which manifests in the smooth GP realizations seen in the left panels, irrespective of the length scale. The functions allowed by the exponential kernel are non-differentiable, which manifests itself in the spiky function realizations seen in the right panels.}
    \label{fig:samp}
\end{figure*}
\begin{figure*}
    \centering
    \includegraphics[width=\linewidth]{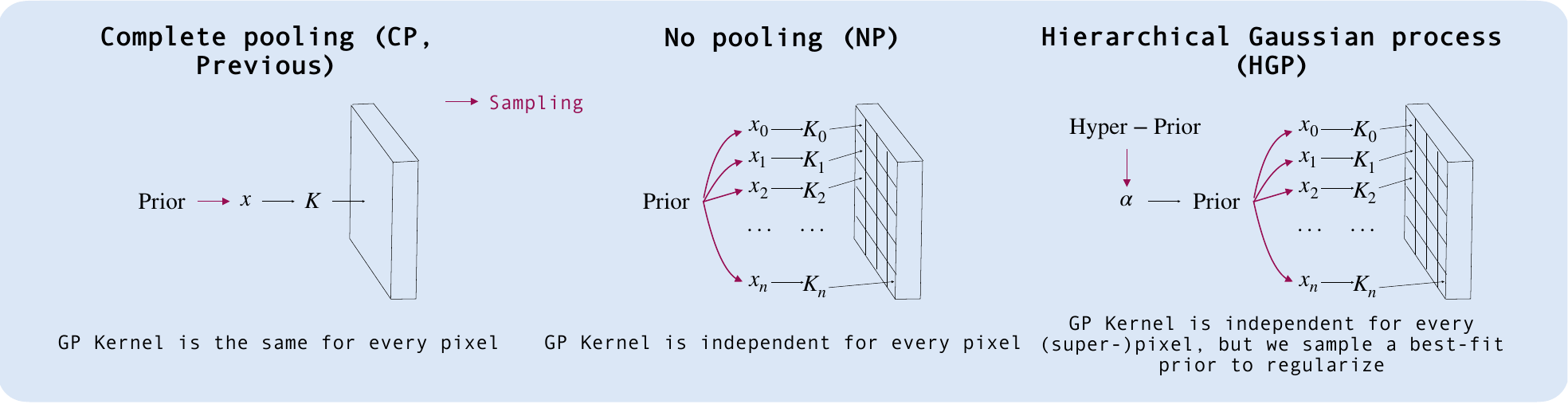}
    \caption{An illustration of different GP model variants. In the CP model, which has been used extensively in previous 21~cm component separation analyses, we sample a set of global kernel parameters for all LoS. For the NP model, we sample independent kernel parameters for every LoS. In the HGP model, the full data cube is divided into a set of superpixels, with kernel parameters being shared amongst all LoS within a superpixel. The kernel hyper-parameters are drawn from a learned prior, which is inferred through the use of a global hyper-prior over the prior parameters. This hyper-prior acts to regularize spatial variations in the kernel parameters and allows the sharing of information between different LoS.}
    \label{fig:gpmv}
\end{figure*}

{In this section, we briefly summarize the GP methodology for \HI{} IM component separation and the GP model variants we adopted in this work. We also provide a general overview of modeling and inference for GPs in Appendix \ref{subsec: gp overview}}.

\subsection{Gaussian Processes for Component Separation}
\label{sec:comp sepa}
\subsubsection{Modeling the sky as GP realizations}

In this paper, we focus on using GP regression for \HI{} IM component separation, and as such consider a sky signal consisting of astrophysical foreground emission, polarization leakage, and the cosmological \HI{} signal. These will be described in detail in Section \ref{sec: sims}. The core idea behind using GP models for component separation is to model the total sky signal as an additive GP realization, i.e., foreground, polarization leakage and \HI{} terms are described {by separate} GP realizations which are each assigned their own GP kernel. {The GP priors and the kernels used in this work are defined below}. Each emission component then can be separated by exploiting the fact that they each display different behavior as a function of frequency, which can be modeled through their GP covariance kernels.

Given an observed data vector $\bi{d}_\mathrm{obs}(\bi{\nu})$ at some $N$ frequencies $\bi{\nu}$, we can model this as a GP realization. A GP assigns a prior over the space of functions $f\sim\mathcal{GP}(m, \kappa)$, where $m$ is some mean function around which GP realizations are distributed and $\kappa$ is the kernel function which describes how function realizations vary around the mean function. The mean function is often set to the zero function, with the GP kernel being the key quantity describing the spectral behavior of the function realizations. For our setting, where we have observations over a set of frequencies, the GP model amounts to describing the observed signal as a draw from the multivariate Gaussian $\bi{f}(\bi{\nu})\sim\mathcal{N}(0, K\equiv \kappa(\bi{\nu},\bi{\nu'}))$, where $K\in\mathbb{R}^{N\times N}$ is a covariance matrix obtained by evaluating the kernel function for each frequency pair.

For our sky model, we follow \citetalias{Soares_2022}, decomposing the sky emission into a cosmological \HI{} component, $\bi{f}_{\rm HI}$ and a foreground component, $\bi{f}_{\rm fg}$. The observed sky signal along a given LoS may then be written as
\begin{equation}
    {\bi{d}_{\rm obs}(\bi{\nu})=\bi{f}_{\rm HI}(\bi{\nu}) + \bi{f}_{\rm fg}(\bi{\nu}) + \bi{\eta}},
    \label{eqn:sky forward problem}
\end{equation}
where ${\bi{\eta}\sim\mathcal{N}(0, \Sigma_\eta)}$ is a Gaussian noise vector with covariance $\Sigma_\eta\in\mathbb{R}^{N\times N}$, {which can be modeled as an extra GP realization. For brevity, in the remainder of this section we drop the explicit frequency dependence in the GP realizations.} The foreground signal is further decomposed as 
\begin{equation}
    {\bi{f}_{\rm fg} = \bi{f}_{\rm sm} + \bi{f}_{\rm pol}},
    \label{eqn:fg signal decomp}
\end{equation}
where {$\bi{f}_{\rm sm}$} is the signal due to astrophysical emission that varies smoothly with frequency, and {$\bi{f}_{\rm pol}$} is the signal due to polarization leakage, {where the signal from polarized emission leaks into the Stokes $I$, total intensity signal due to imperfect antenna alignments \citep[see e.g. Section 6.3 of][]{2020PASP..132f2001L}}. Modeling each of these terms as GP realizations amounts to assuming,
\begin{equation}
\begin{aligned}
    {\bi{f}_{\rm sm}\sim \mathcal{N}(0,K_{\rm sm})},\\
    {\bi{f}_{\rm pol}\sim \mathcal{N}(0,K_{\rm pol})},\\
    {\bi{f}_{\rm HI}\sim \mathcal{N}(0,K_{\rm HI})},
\end{aligned}  
\end{equation}
where $K_{\rm HI}$, $K_{\rm sm}$, and $K_{\rm pol}$ are the GP covariances for the HI, smooth astrophysical foreground, and polarization leakage terms respectively.

{Assuming no cross-correlation between different emission components, we may use the fact that the covariance of the sum of independent Gaussian random variables is the sum of the covariances to write} the total GP covariance,
\begin{equation}
    K=K_{\rm HI}+K_{\rm fg}+{\sigma_\eta^2 I_N}=K_{\rm HI} + K_{\rm sm} + K_{\rm pol}+{\sigma_\eta^2 I_N}.
    \label{eqn: full GP kernel}
\end{equation}
{We assume a fixed noise variance for all frequencies such that $\Sigma_\nu=\sigma_\eta^2 I_N$, where $I_N$ is the $N\times N$ identity matrix}. {For this particular case, where we have assumed Gaussian noise, we then have $\bi{d}_{\rm obs}\sim\mathcal{N}(0,K)$}.

\subsubsection{GP Component Separation}
\label{sec: GP sep}
From the observation model defined through Equation~(\ref{eqn:sky forward problem}) {and GP modeling for each of the components, the joint distribution of {$(\bi{f}_{\rm fg},\bi{d}_{\rm obs})$} can be written as}
\begin{equation}
    {\begin{bmatrix} \bi{f}_{\rm fg}  \\ \bi{d}_{\rm obs} \end{bmatrix} \sim \mathcal{N} \left( \mathbf{0}, \begin{bmatrix} K_{\rm fg} & K_{\rm fg} \\ K_{\rm fg} & K \end{bmatrix} \right).}
\end{equation}
{The covariance between $\bi{f}_{\rm fg}$ and $\bi{d}_{\rm obs}$ is given by $K_{\rm fg}$ because the cross-correlation between $\bi{f}_{\rm fg}$ and the other components in $\bi{d}_{\rm obs}$ is zero.}

We can then write down the prediction formulae for the expectation and covariance of the foreground signal {given a fixed GP covariance kernel $K$ and observed signal $\bi{d}_{\rm obs}$},
\begin{equation}
\begin{aligned}
    \mathbb{E}\left[{\bi{f}_{\rm fg}}|K,{\bi{d}_{\rm obs}}\right]&=K_{\rm fg}(K_{\rm fg}+K_{\rm HI}+\sigma_\eta^2 I_N)^{-1}{\bi{d}_{\rm obs}},\\
    \mathrm{cov}\left[{\bi{f}_{\rm fg}}|K\right]&=K_{\rm fg}-K_{\rm fg}\left(K_{\rm fg}+K_{\rm HI}+\sigma_\eta^2 I_N\right)^{-1}K_{\rm fg},
    \label{eqn:foreground prediction formulae}
\end{aligned}
\end{equation}
Given predictions for the foreground signal, $\mathbb{E}[{\bi{f}_{\rm fg}}| K,{\bi{d}_{\rm obs}}]$, one can calculate the residual ${\bi{f}_{\rm res}}$, {which consists of our estimate of the 21~cm signal and instrumental noise},
\begin{equation}
    {\bi{f}_{\rm res} = \bi{d}_{\rm obs} - \mathbb{E}[\bi{f}_{\rm fg}|K,\bi{d}_{\rm obs}]}.
    \label{eqn:21cm est}
\end{equation}

\subsubsection{Sampling the GP Posterior}

{In order to compute the expected foreground signal we must be able to estimate the GP covariance $K$. For the Gaussian noise model we assume in this work, we can analytically marginalize over the GP realizations, giving the log-marginal likelihood
\begin{equation}
    {\log \mathcal{L}(\bi{d}_{\rm obs}|K)=-\frac{1}{2}\bi{d}_{\rm obs}^TK^{-1}\bi{d}_{\rm obs}-\frac{1}{2}\log|K|-\frac{N}{2}\log2\pi}.
    \label{eqn: lml}
\end{equation}}
{In general, we may parameterize the GP covariance through some set of parameters $\Phi$, to which we assign the prior $\pi_\Phi(\Phi)$. Our goal is then to infer the values of these parameters $\Phi$. The log-marginal posterior is given by
\begin{equation}
    {\log p(\Phi|\bi{d}_{\rm obs}) = \log \mathcal{L}(\bi{d}_{\rm obs}|\Phi)+\log \pi(\Phi)},
\end{equation}
where we have explicitly conditioned the log-marginal likelihood on the covariance kernel parameters $\Phi$.}

A typical approach at this point is to optimize the log-marginal posterior with respect to $\Phi$, then fixing their values at the maximum a-posteriori (MAP) values for the downstream inference tasks such as component separation in this work \citep{RasmussenW06}. This can be viewed under the empirical Bayes framework, where the data are used to set the prior parameters, which are fixed for inference over the remaining model parameters. This approach has the advantage of being fast compared to full integration over the marginal posterior. However, it comes with two significant drawbacks. By fixing $\Phi$ at their MAP values we do not fully propagate the posterior uncertainty on them to our inferences over {$\mathbb{E}[\bi{f}_{\rm fg}|\bi{d}_{\rm obs}]$}. In addition, the MAP estimate for parameter values is generally not a good estimator. For many models, using the MAP value as an estimator can lead to poor predictive performance and can result in significantly biased parameter inferences \citep{mackay2003information, betancourt2017conceptual}. This has previously been noted in GP models, where MAP estimation has been observed to tend to result in overfitting \citep{lalchand2020approximate}. Problems with MAP estimation are particularly apparent for hierarchical Bayesian models, which we consider in this work, where correlations between model hyperparameters and latent variables result in posterior geometries where the MAP is very far from the typical set, i.e., where the probability mass of the posterior is concentrated \citep{betancourt2015hamiltonian, betancourt2017conceptual, hoffman2019neutra}. 

Given these problems with MAP estimation, we consider fully Bayesian GP models throughout this work, and demonstrate the degradation in our cosmological signal recovery using MAP estimation in Section \ref{subsec:MAP vs sampling}. The final expected foreground and 21~cm signals, along with their corresponding uncertainties, can be obtained {by calculating the ensemble expectations over the GP covariance parameter posterior samples $\{\Phi^j\}_{j=1}^{M}$,} 
\begin{equation}
    {\mathbb{E}\left[\bi{f}_{\rm fg}|\bi{d}_{\rm obs}\right] = \frac{1}{M}\sum_{j=1}^{M}\mathbb{E}\left[\bi{f}_{\rm fg}|\Phi^j,\bi{d}_{\rm obs}\right],}
\end{equation}
\begin{equation}
\begin{aligned}
{\mathrm{cov}\left[\bi{f}_{\rm fg}|\bi{d}_{\rm obs}\right]=\frac{1}{M-1}\sum_{j=1}^{M}\left\{\left(\bi{f}_{\rm fg, pp}^j - \mathbb{E}[\bi{f}_{\rm fg}|\bi{d}_{\rm obs}]\right)\right.}\\{ \left.\otimes \left(\bi{f}_{\rm fg, pp}^j - \mathbb{E}[\bi{f}_{\rm fg}|\bi{d}_{\rm obs}]\right)\right\}.}
\end{aligned}
\end{equation}
{The covariance is estimated using the posterior predictive samples, $f_{\rm fg, pp}^j\sim\mathcal{N}(\mathbb{E}\left[\bi{f}_{\rm fg}|\Phi^j,\bi{d}_{\rm obs}\right], \mathrm{cov}[\bi{f}_{\rm fg}|\Phi^j])$, which can be obtained using the prediction formulae in Equation (\ref{eqn:foreground prediction formulae}), for each of the posterior samples $\Phi^j$.}

\subsubsection{Emission Component Covariance Kernels}

We follow \citetalias{Soares_2022} in selecting the kernels for our emission components. Astrophysical foreground emission is expected to vary smoothly as a function of frequency. As such, astrophysical foregrounds are modeled through a radial basis function (RBF) kernel,
\begin{equation}
    {\kappa}_{\rm sm}(\nu,\nu^\prime) = \sigma^2_{\rm sm}\exp{\left(-\frac{\lVert\nu-\nu^\prime\lVert^2}{2l_{\rm sm}^2}\right)},
    \label{eqn:fgkernel}
\end{equation}
where $\sigma_{\rm sm}$ is a hyperparameter that controls the amplitude of the spectrum, whilst $l_{\rm sm}$ determines the correlation between frequency channels. Larger values of $l_{\rm sm}$ indicate data in all frequency channels are highly correlated, resulting in a smoother spectrum. The left panel of Figure \ref{fig:samp} shows samples from a GP with an RBF kernel, with kernel variance $\sigma^2=1$ and length scales $l=0.5$ and $l=0.05$. It can be seen that the GP samples vary smoothly for both length scales, which is to be expected given that the functions allowed by the RBF kernel are infinitely differentiable.

For the 21~cm signal, we apply the exponential kernel,
\begin{equation}
    {\kappa}_{\rm HI}(\nu,\nu^\prime) = \sigma^2_{\rm HI}\exp{\left( -\frac{\lVert\nu-\nu^\prime\rVert}{2l_{\rm HI}}\right)}.
    \label{eqn:hikernel}
\end{equation}
The exponential kernel corresponds to the Matern kernel with $\rho=1/2$. The functions allowed by this kernel are non-differentiable, and therefore vary in a very spiky fashion with frequency, as demonstrated in the right panel of Figure \ref{fig:samp}. The $\sigma_{\rm HI}$ and $l_{\rm HI}$ parameters control the amplitude and correlation length scale of the \HI{} signal.

Alongside astrophysical foreground emission, we also consider the effects of polarization leakage in our simulations. This is also expected to vary smoothly with frequency and is therefore modeled with an RBF kernel,
\begin{equation}
    {\kappa}_{\rm pol}(\nu,\nu^\prime) = \sigma^2_{\rm pol}\exp{\left(-\frac{\lVert\nu-\nu^\prime\rVert^2}{2l_{\rm pol}^2}\right)},
    \label{eqn:polkernel}
\end{equation}
where $\sigma_{\rm pol}$ and $l_{\rm pol}$ control the amplitude and correlation length scale of the polarization leakage respectively. In this work we will consider a three-kernel model, with the total GP kernel being given by Equation~(\ref{eqn: full GP kernel}), and also a two-kernel model where we model the emission due to astrophysical foregrounds and polarization leakage through a single RBF kernel. For the two-kernel model, the total GP kernel can be written as $K=K_{\rm HI} + K_{\rm sm}$.

\subsection{Gaussian Process Model Variants}
\label{sec:gpmodels}
In this section, we summarize the different GP models that we study in this work, which account for spatial variations in foreground emission in different ways. A graphical illustration of the models is shown in Figure \ref{fig:gpmv}.

\subsubsection{Complete Pooling (CP) model}
The standard GP model used for component separation (e.g., \citetalias{Soares_2022}) uses the same set of kernels to describe the emission spectrum along every LoS, i.e., the kernel parameters for every LoS are pooled to a single value. In this work, we refer to this as the complete pooling (CP) model, which is used as our baseline throughout.

For the GP kernel parameters, we use a set of weakly informative priors, chosen to place the probability mass over the expected scales of the parameters. In general, when information on the expected scale of a parameter is available, it is appropriate to select a prior distribution that concentrates its probability mass over those scales. For example, for a strictly positive parameter $\theta$ that is expected to have a value $\sim 1$, one might select the half-normal prior $\theta\sim\rm{HalfNormal}(\sigma=1)$. This has the benefit of helping to regularize our inferences using prior information and avoids the pitfalls associated with common choices for so-called uninformative prior distributions. For example, selecting a very broad uniform distribution as a default prior concentrates most of the prior probability mass at extreme values within the support of the distribution, which can bias the subsequent posterior inference. From an algorithmic perspective, this can also cause problems for Markov Chain Monte Carlo (MCMC) algorithms. Without regularization from a weakly informative prior, the posterior geometry can be such that it severely frustrates convergence, in some cases jeopardizing the geometric ergodicity of the MCMC algorithm. These issues appear frequently in the fully Bayesian GP and hierarchical models we consider in this work. For a detailed discussion of prior choice and the impacts on inference and sampling, we refer the reader to \citep{gelman2013bayesian, stan_development_team_stan_2012, betancourt2015hamiltonian, gelman2017beyond, 2017Entrp..19..555G, gabry2019visualization}.

\begin{table}
    \centering
    \begin{tabular}{|c|c|}
        \hline
         Parameter& Prior\\ 
         \hline
         $\sigma_{\rm sm}^2/10^{-2}$& $\rm{LogNormal}(0,4)$\\ 
         $l_{\rm sm}$& $\rm{InverseGamma}(2,1)$\\ 
         $\sigma_{\rm pol}^2/10^{-6}$& $\rm{LogNormal}(0,4)$\\ 
         $l_{\rm pol}$& $\rm{InverseGamma}(5,1)$\\ 
         $\sigma_{\rm HI}^2/10^{-8}$& $\rm{HalfNormal}(1)$\\ 
         $l_{\rm HI}$& $\rm{HalfNormal}(0.02)$\\ 
 $\sigma_\eta/10^{-7}$&$\rm{HalfNormal}(10)$\\ 
 \hline
    \end{tabular}
    \caption{Priors for the CP model. Here $\rm{LogNormal}(\mu, \varsigma)$ corresponds to a log-normal distribution with mean parameter $\mu$ and scale parameter $\varsigma$, $\rm{InverseGamma}(\alpha, \beta)$ is the inverse-gamma distribution with shape parameter $\alpha$ and scale parameter $\beta$, and $\rm{HalfNormal}(\varsigma)$ is the half-normal distribution with scale parameter $\varsigma$. Note that for a two-kernel model we exclude the polarization leakage parameters, $\sigma_{\rm pol}$ and $l_{\rm pol}$.}
    \label{tab:CP priors}
\end{table}

The weakly informative priors for the GP kernel parameters in the CP model are stated in Table \ref{tab:CP priors}. The astrophysical foreground and polarization leakage components are expected to be orders of magnitude brighter than the \HI{} signal. For the GP variance parameters of these components, we select zero-avoiding log-normal priors that concentrate over the expected scales of the two emission components. For the length-scale parameters of the foreground components, we use inverse-gamma priors. The inverse-gamma distribution is a common choice for GP length scale priors. The distribution places negligible probability masses on very small length scales. For length scales below the minimum covariate spacing, the GP likelihood will plateau, putting considerable probability mass at these small scales. This can result in overfitting and also induce convergence problems for the MCMC algorithms \citep{stan_development_team_stan_2012}. For the smoothly varying foreground emission components we do not expect such high-frequency signals, and therefore use the inverse-gamma prior to remove these modes. The remainder of the prior concentrates around the expected length scales of the foreground components, whilst having a heavy right tail to allow for low-frequency modes. 

For the \HI{} signal, we choose a half-normal prior for the GP kernel variance. This allows for potentially zero cosmological signal whilst down-weighting very large amplitude scales. For the length scale parameter, in this case, we also use a half normal prior, allowing the \HI{} signal to be potentially completely uncorrelated in frequency, and marginalizing over any very high-frequency modes. We find that using a half-normal prior for the faint \HI{} component does not frustrate the convergence of our sampling algorithm.

We account for noise in our GP model by fitting for a single noise standard deviation $\sigma_\eta$, to which we assign a half-normal prior. In real experiments, the true noise covariance can be challenging to estimate. This can be accounted for by marginalizing over the noise process, with some prior selected to concentrate over the expected noise scales. In more realistic setups, we may expect more complicated noise contributions e.g., heteroscedastic noise, where it would be appropriate to use a more sophisticated noise model than the uniform white noise we consider here. We leave more realistic noise modeling to future work, focusing here on spatial variations in foreground emission. We use the same noise model for all the GP models considered in this work.

\subsubsection{No-Pooling (NP) model}

The CP model assumes that the GP kernel parameters for the foreground emission are identical for every LoS. However, the spectral properties of diffuse Galactic emission are known to vary significantly across the sky \citep{2016A&A...594A..10P, 2018JCAP...04..023R, 2020MNRAS.495..578J}. Using a single set of GP kernel parameters for every LoS may fail to capture these variations, potentially resulting in mismodeling and a failure to accurately recover the underlying sky emission. At the opposite extreme from the CP model, we may consider a no-pooling (NP) model, where the foreground GP kernel parameters are assumed to vary completely independently along every LoS. We do not expect such variations in the \HI{} GP prior along each LoS, and as such we retain the global kernel for the \HI{} signal used in the CP model. The priors for the GP kernel parameters take the same form as in Table \ref{tab:CP priors}, with the difference that now we have separate $\{\sigma_{\rm{sm},i}, l_{\rm{sm}, i}, \sigma_{\rm{pol}, i}, l_{\rm{pol}, i}\}$ for each LoS $i$.

\subsubsection{Hierarchical Gaussian Process (HGP) model}

Sitting between the CP and NP models described above, we may consider a class of hierarchical Gaussian process (HGP) models. In this setting, we divide the whole data set into a set of superpixels containing multiple LoS, where the superpixel size is smaller than the full data set over the spatial dimensions. Within each superpixel, the smooth foreground and polarization leakage kernel parameters are assumed to be constant. Ideally, we would consider separate kernel parameters for each LoS. However, the memory requirements for such a setup render such a model intractable given our computational resources. We discuss the details of the HGP model setup in Section \ref{subsec: gp sep setup}.  
\begin{table}
    \centering
    \begin{tabular}{|c|c|}
        \hline
         Parameter& Prior\\
         \hline
         $\varsigma_{\rm sm}$& $\rm{LogNormal}(0,4)$\\
         $\alpha_{\rm sm}$& $\rm{LogNormal}(1,4)$\\
         $\beta_{\rm sm}$& $\rm{LogNormal}(0,4)$\\
         $\varsigma_{\rm pol}$& $\rm{LogNormal}(0,4)$\\
         $\alpha_{\rm pol}$& $\rm{LogNormal}(1,4)$\\
         $\beta_{\rm pol}$& $\rm{LogNormal}(0,4)$\\
         $\sigma_{\mathrm{sm}, p}^2/10^{-2}$& $\rm{HalfNormal}(\varsigma_{\rm sm})$\\
         $l_{\mathrm{sm}, p}$& $\rm{InverseGamma}(\alpha_{\rm sm},\beta_{\rm sm})$\\
         $\sigma_{\mathrm{pol}, p}^2/10^{-6}$& $\rm{HalfNormal}(\varsigma_{\rm pol})$\\
 $l_{\mathrm{pol}, p}$&$\rm{InverseGamma}(\alpha_{\rm pol},\beta_{\rm pol})$\\
 $\sigma_{\rm HI}^2/10^{-8}$&$\rm{HalfNormal}(1)$\\
 $l_{\rm HI}$&$\rm{HalfNormal}(0.02)$\\
 $\sigma_\eta/10^{-7}$&$\rm{HalfNormal}(10)$\\
 \hline
    \end{tabular}
    \caption{Priors for the HGP model, where the subscript $p$ denotes the superpixel index. For the two-kernel model we exclude the polarization leakage parameters, $\varsigma_{\rm pol}$, $\alpha_{\rm pol}$, $\sigma_{\rm pol}$ and $l_{\rm pol}$.}
    \label{tab:HGP priors}
\end{table}
The priors for the HGP model are given in Table \ref{tab:HGP priors}. The priors for the foreground kernel parameters in a given superpixel $p$ depend on a set of global hyper-parameters $\{\varsigma_{\rm sm}, \alpha_{\rm sm}, \beta_{\rm sm}, \varsigma_{\rm pol}, \alpha_{\rm pol}, \beta_{\rm pol}\}$. To these global hyper-parameters, we assign a set of weakly informative hyper-priors, which are chosen such that the potential priors over the foreground kernel parameters for each superpixel correspond broadly to the expected scales of those foreground parameters. A key feature of the hierarchical approach is that we jointly fit for these hyper-parameters, allowing us to learn the effective prior over the superpixel parameters and share information between superpixels. This allows for spatial variations in the kernel parameters, whilst regularizing those variations through the hyper-prior. This can help to reduce the tendency for NP models to overfit, particularly with very noisy data where the assumption of total independence means that there is no regularizing effect from the global hyperprior. As in the NP model, we assume a global \HI{} kernel.

\section{The 21~cm component separation simulations and computation}\label{sec: sims}

In this section, we describe the simulated data set used to test our GP models for 21~cm component separation, and the computational methods and setup we use for inference.

\subsection{Mock data set}
{The} simulated sky signal consists of foreground emission, instrumental effects, and a cosmological 21~cm signal. We adopted the data set\footnote{{A description of this data set can be found in \href{https://github.com/paulassoares/gpr4im/tree/main/Data}{this repository}, while the data set is provided at \href{https://www.dropbox.com/sh/9zftczeypu7xgt3/AABiiBw_0SBPrLgSHsjiISz8a?dl=0}{this dropbox link}}.} provided by \citetalias{Soares_2022}, which was originally used in \citet{Cunnington2021}. {The original data set has a spatial dimension of (256, 256) pixels and 285 frequency channels, corresponding to a redshift range of $0.2<z<0.58$. The physical size of the data cube is $(L_x,L_y,L_z)=(1436,1436,1340)\: \rm Mpc$, and the frequency resolution of the simulated data set is $\delta \nu = 1\:\rm MHz$. At the central redshift $z=0.39$, this physical size corresponds to a sky area of 2927 deg$^2$. We cut 256 frequency channels from the total data cube, resulting in a grid of size $(N_x,N_y,N_z)=(256,256,256)$. The corresponding box size is $(L_x,L_y,L_z)=(1436,1436,1193)\: \rm Mpc$, with a frequency range along the $z$-axis of $899\: \rm{MHz}<\nu<1155\: \rm{MHz}$, corresponding to a redshift range of $0.23<z<0.58$.}


The simulated data serve as a conceptual model for \HI{} IM surveys, akin to the MeerKLASS survey \citep[e.g.,][]{2016mks..confE..32S,2022MNRAS.509.4923I, 2024arXiv240721626M}. A flat-sky approximation is applied here, with curved-sky effects over the small survey area being negligible. In the following we summarize the simulation procedure for each of the sky components, referring the readers to Appendix A in \citetalias{Soares_2022} and \citet{Cunnington2021} for further details.

\subsubsection{Foreground Model}

The total foreground emission, $\delta T_{\rm FG}$ consists of four components,
\begin{equation}
    \delta T_{\rm FG} = \delta T_{\rm sync} +\delta T_{\rm free} +\delta T_{\rm point} +\delta T_{\rm pol}, 
\end{equation}
where $\delta T_{\rm sync}$ is due to Galactic synchrotron emission, $\delta T_{\rm free}$ is due to Galactic free-free emission, $\delta T_{\rm point}$ is due to extragalactic point sources, and $\delta T_{\rm pol}$ is due to polarization leakage.

Galactic synchrotron emission is caused by cosmic-ray electrons spiraling in the Galactic magnetic field. The Planck Legacy Archive\footnote{\url{http://pla.esac.esa.int/pla/}} FFP10 simulations were used to generate synchrotron emission maps at 217 and 353 GHz. These simulations used the 408 MHz all-sky map with $56\ \rm arcmin$ resolution as a synchrotron emission template, enhanced to a Healpix \citep{2005ApJ...622..759G} resolution of $N_{\rm side} = 2048$ using Gaussian random field realizations \citep{1981A&A...100..209H, 1982A&AS...47....1H, 2015MNRAS.451.4311R}. A spectral index map can be derived from the simulated 217 and 353 GHz maps, assuming a power-law emission model, which can then be used to extrapolate our simulated map over the observational frequency range.

Galactic free-free emission is caused by free electrons scattering off ions in an ionized gas. The simulated data set uses the 217 GHz FFP10 free-free simulation at $N_{\rm side} = 2048$. This free-free map was generated using templates from \cite{2003MNRAS.341..369D}, and WMAP Maximum Entropy Method (MEM) derived free-free maps. The simulated map is extrapolated over the observed frequency range using a power-law emission model with a constant spectral index over all pixels.

Extragalactic point sources, such as Active Galactic Nuclei, are included by applying a polynomial fit to 1.4 GHz radio sources as in \cite{2013MNRAS.434.1239B}, and scaling to the target observational frequencies using the method described in \cite{2018MNRAS.473.4242O}.

Polarization leakage is caused by {mis-alignment of telescope antennas, which leads to} some fraction of the {linearly polarized} synchrotron Stokes $Q$ and $U$ leaking into Stokes $I$. {As synchrotron signals propagate through the Galactic interstellar medium they undergo Faraday rotation, where the polarization angle is rotated. This Faraday rotation is frequency dependent, which induces a frequency dependence in the polarization leakage, and hence} additional structure in the synchrotron emission spectrum, complicating GP-based component separation methods that exploit the spectral smoothness of the foreground emission. {Efforts have been made to model \citep[e.g.][]{2015PhRvD..91h3514S, 2014MNRAS.444.3183A} and control \citep[e.g.][]{2001A&A...375..344B,2017ApJ...848...47N} this effect.}
{The} polarization leakage {signal used in this work} was simulated using the \textsc{CRIME}\footnote{\url{http://intensitymapping.physics.ox.ac.uk/CRIME.html}} software package. This was used to produce Stokes $Q$ emission maps over the observational frequency range, subsequently assuming a polarization leakage level of 0.5\% from Stokes $Q$, {and zero leakage from Stokes $U$ for simplicity} \citep{2014MNRAS.444.3183A, Cunnington2021}.

\subsubsection{Instrument and Noise Model}

In \citetalias{Soares_2022} a MeerKLASS-like \citep{2016mks..confE..32S} survey is simulated. The corresponding observational noise is assumed to be uncorrelated in frequency, and drawn from a Gaussian distribution with standard deviation at each frequency given by
\begin{equation}
    \sigma_\eta(\nu)=T_{\rm sys}(\nu)\left(\delta_\nu t_{\rm tot}\frac{\Omega_p N_{\rm dish}}{\Omega_a }\right)^{-1/2},
\end{equation}
where $T_{\rm sys}(\nu)$ is the frequency-dependent system temperature, {which is set to $T_{\rm sys}(\nu)[{\rm K}] = 25 + 66\times (300/\nu[{\rm MHz}])^{2.55}$}, $\delta_\nu=1\,\rm{MHz}$ is the frequency resolution of the observations, $t_{\rm tot}=1000\:\rm{hours}$ is the total observing time, {$\Omega_p = 1.13\theta_{\rm FWHM}^2$} is the pixel solid angle, {$\theta_{\rm FWHM}$ is the beam full width at half maximum (FWHM)}, {$\Omega_a \approx 0.89\,\mathrm{sr}$} is the survey solid angle and $N_{\rm dish}=64$ is the number of dishes (identical to the MeerKAT {telescope}). The noise root-mean-square (RMS) in the simulation ranges from $0.032$ to $0.040\: \rm mK$ over the observed frequency range.

The total sky signal is smoothed with a constant Gaussian beam corresponding to the angular resolution of the instrument. The beam FWHM is given by
\begin{equation}
    \theta_{\rm FWHM} = \frac{1.22c}{\nu D_{\rm dish}},
\end{equation}
where $c$ is the speed of light, $\nu$ is the observed frequency {(taken to be 899 MHz)}, and $D_{\rm dish}$ is the telescope dish diameter. The telescope dish diameter is set to {$D_{\rm dish}=15\,\mathrm{m}$}, consistent with SKA1-Mid dishes \citep{dewdney2019ska1} {and close to the MeerKAT dish size of 13.5m}.

\subsubsection{Cosmological Signal}

{For the simulations we adopted from \citetalias{Soares_2022}, the cosmological signal was} obtained from the MultiDarkPlanck (MDPL2) N-body simulation \citep{Klypin2016}, which simulates $3840^3$ dark matter particles within a cubic volume with a size of $1\: {\rm Gpc}/h$ on each side, using the cosmological parameters of Planck 2015 \citep{2016A&A...594A..13P}. The MultiDark-SAGE catalog derived from this simulation, available on the Skies \& Universe\footnote{\url{skiesanduniverses.org}} website, was selected for analysis. This simulation provides redshift snapshots, each representing the state of the cosmological density field and galaxies at different redshifts. For the simulated data set used in this work, the snapshot $z = 0.39$ was used, extending an assumed redshift range of $z=0.2$ to $z=0.58$. Galaxies in this snapshot were organized into voxels using the Nearest Grid Point (NGP) assignment. 

The \HI{} mass was calculated from the cold gas mass of each galaxy in the catalog and combined within each voxel to form an \HI{} intensity map. However, this approach excludes halos lighter than $10^{10}h^{-1}M_{\odot}$. To compensate, the {\HI{} mass inside these small halos down to $10^{8}h^{-1}M_{\odot}$ was calculated} based on measurements from the GBTWiggleZ cross-correlation analysis \citep{2013ApJ...763L..20M}, {which gives the abundance of HI in a halo $\Omega_{\rm HI} = [4.3\pm 1.1]\times 10^{-5}/b_{\rm HI}r$. The halo \HI{} bias fit in \cite{2018ApJ...866..135V} gives the \HI{} bias $b_{\rm HI} = 1.105$ at the central redshift $z=0.39$, and the cross-correlation coefficient $r$ is set to 1.} Each redshift slice in the \HI{} IM simulation was then converted into an overtemperature field by subtracting the mean temperature, with the overtemperature field directly tracing the matter overdensity field. {The 21 cm signal field adopted in this work has a standard deviation of $\sim 0.30$ mK, while the noise standard deviation is $\sim 0.035$ mK. Thus, the residual  $\bi{f}_{\rm res}$ is expected to be signal dominated, along with the corresponding residual power spectra.}

\subsection{GP Component Separation Setup}\label{subsec: gp sep setup}

\begin{table}
\centering
\setlength{\tabcolsep}{6mm}{
\begin{tabular}{|l|c|}
\hline
\textbf{GP model} & \textbf{Model description} \\
\textbf{abbreviation} & \\
\hline
CP2\tablenotemark{*} & Complete pooling model \\
    & with 2 kernels \\
CP3\tablenotemark{*} & Complete pooling model\\
    & with 3 kernels \\
HGP2 & Hierarchical model\\
    & with 2 kernels \\
HGP3 & Hierarchical model\\
    & with 3 kernels \\
NP2 & No pooling model\\
    & with 2 kernels \\
NP3 & No pooling model\\
    & with 3 kernels \\
\hline
\end{tabular}}
\tablenotetext{*}{{The CP3 and CP2 models are fully Bayesian implementation of the \citetalias{Soares_2022} GP models with three and two GP kernels respectively.}}
\caption{The model name abbreviations used throughout this work. Models with two kernels include an \HI{} kernel and a single foreground kernel. Models with three kernels include an \HI{} kernel and two foreground kernels, with one foreground kernel modeling smooth astrophysical emission, and the other foreground kernel modeling polarization leakage.}

\label{table:acronyms}

\end{table}

In Table \ref{table:acronyms} we list the model name abbreviations used in this work. {We applied all models to the same data set described in Section \ref{sec: sims} to assess their performance}. As discussed in Section \ref{sec: GP sep}, we consider fully Bayesian GP models throughout this work. This requires us to perform sampling over the full GP hyperparameter posterior. In the case of the NP and HGP models, this is a very high-dimensional sampling problem, which is made tractable by using the No-U-Turn Sampler (NUTS) algorithm \citep{hoffman2014no}. The NUTS algorithm is an adaptive variant of the Hamiltonian Monte Carlo (HMC) algorithm. HMC exploits gradient information of the target posterior to generate efficient sampling proposals by following Hamiltonian trajectories through the target phase space and has excellent scaling to high dimensions. Our code can be found in \href{https://github.com/dkn16/H21F}{this repository}.

For each GP model, we use NUTS to sample the parameters {with one chain, running 1000 iterations for burn-in, the samples from which are discarded, followed by 2000 sampling iterations}. Convergence is checked by analyzing the Gelman-Rubin $\hat{R}$ statistic \citep{1992StaSc...7..457G}, which compares the between-chain and within-chain variance. {In our one-chain case, the chain is divided into 4 sub-chains to calculate $\hat{R}$}. Values of $\hat{R}\approx 1$ are indicative of convergence, and in this work a strict requirement of $\hat{R}\leq 1.01$ is used. In addition to this, we also monitor for divergences during sampling. Divergences occur when the value of the Hamiltonian diverges whilst generating trajectories through the target phase space. This can be indicative of regions of the posterior where the curvature is such that it cannot be resolved by the sampler, and it can imply failures in geometric ergodicity \citep{betancourt2015hamiltonian, livingstone_gergo2016}. Such issues are common in fully Bayesian GP and hierarchical models. We therefore ensure that zero divergences occur during sampling. Our framework is built with \texttt{numpyro}\footnote{\url{https://github.com/pyro-ppl/numpyro}} \citep{phan2019composable,bingham2019pyro}, a probabilistic programming language built on \texttt{JAX}\footnote{\url{https://jax.readthedocs.io}}. After we obtain the desired hyperparameter samples, we thin the chain to 50 samples. For each sample, we recover the expected {$\bi{f}_{\rm res}$} signal using Equation~(\ref{eqn:21cm est}). Our final estimated {$\bi{f}_{\rm res}$} signal cube is calculated from the ensemble mean of the 50 recovered {$\bi{f}_{\rm res}$} cubes.

For the CP model, we follow the setup described in \citetalias{Soares_2022}. For the NP model, although the foreground GP kernel parameters should be independent for every LoS, the assumption of a global \HI{} kernel requires the full data cube to be fit simultaneously. The memory requirements of the NP model in this case render it intractable, demanding approximately 500 GB in memory. Therefore, to reduce the memory requirements of the NP model, we split the whole data set into 64 subsets of the size $(32,32,256)$. We run 64 NP models on each subset separately, with the \HI{} kernel being shared across every $32{\rm pix}\times 32{\rm pix}$ LoS in each subset. For each data subset, we have $\sim 4\times 10^3$ hyperparameters to sample for the NP3 model, a task that is rendered tractable through the use of the NUTS algorithm. Whilst we would ideally allow the foreground GP kernel parameters to vary along every LoS in the HGP models, the memory requirements of this approach also render such a model intractable using the exact GPs we consider in this work. For the HGP models, we therefore grid the data set into a set of superpixels along the LoS axis. The foreground GP kernel parameters are assumed to be constant for every LoS within a superpixel, with the \HI{} kernel treated as global in the entire data cube. The default size of a single superpixel is set to be $16 \times 16$ normal pixels, a trade-off between the accuracy and computational cost of the model (in both memory and clock time) in our experiments. The performance of HGP models with different superpixel sizes is compared in Section \ref{subsec:super-pixel}.



\begin{figure*}
    \centering
    \includegraphics[width=\linewidth]{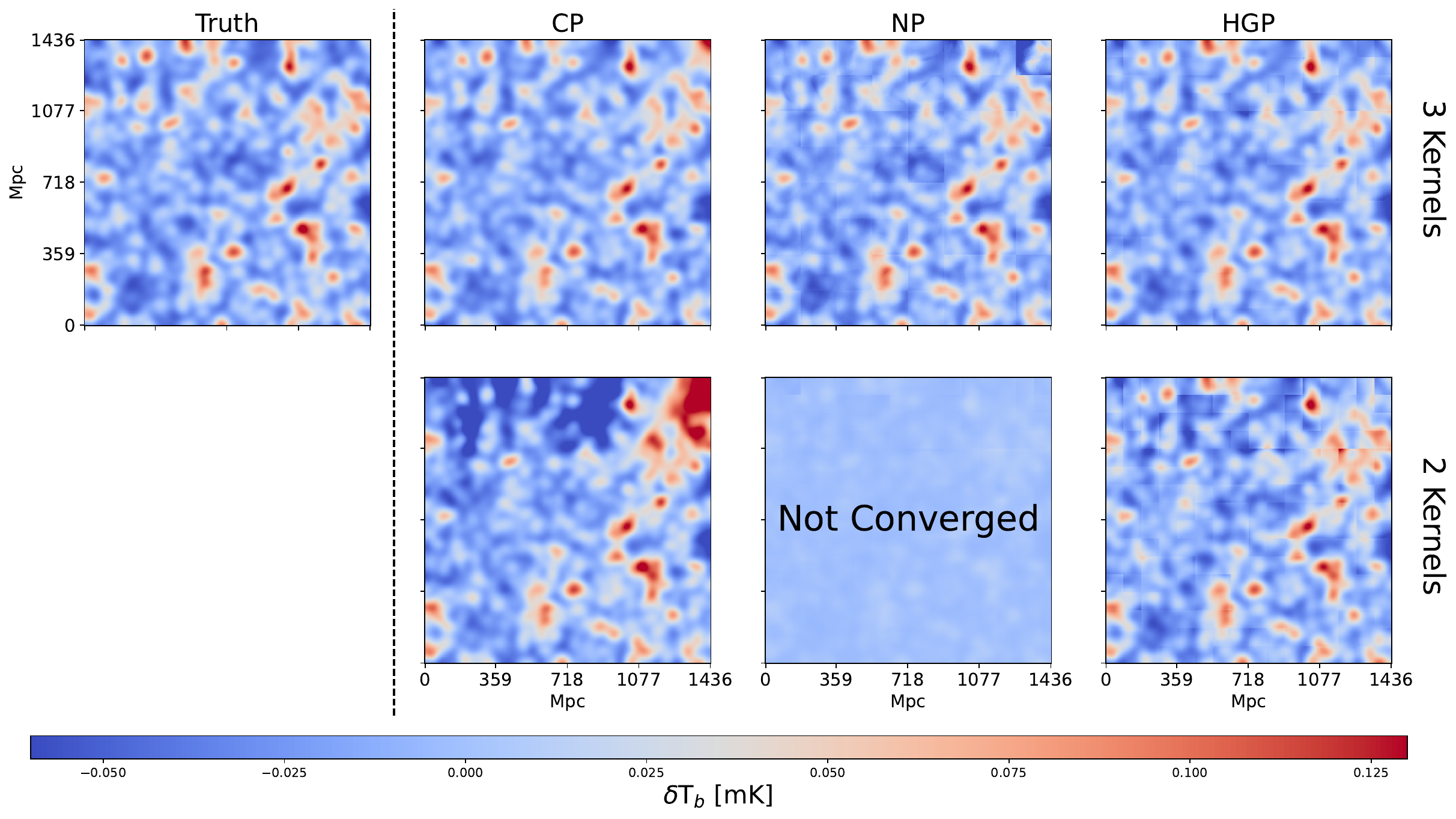}
    \caption{A visual comparison of the recovered {$\bi{f}_{\rm res}$} for the different models. Here we show {2D slices of the $\bi{f}_{\rm res}$ $T_{b}$} at 1019 MHz. The leftmost figure shows the true {$\bi{f}_{\rm res}$} field, separated from the recovered fields by a dashed line. Proceeding from left to right we show the recovered fields for the CP, NP, and HGP models respectively. The top row shows the recovered fields for the three kernel models, and the bottom row shows the recovered fields for the two kernel models. The CP3 and NP3 models have visible foreground residuals in the top right of the field, but these residuals are removed by the HGP3 model. The CP2 model gives poor recovery over much of the observed field, and the NP2 model failed to converge. The HGP2 model obtains similar residuals to the CP3 model. Whilst the NP3, HGP3, and HGP2 models have visible boundary effects from the use of data subsets and superpixels, the models still achieve improved recovery compared to the CP3 model as measured through metrics such as predictive performance (for the HGP3 and HGP2 models), and the recovery of summary statistics.}
    \label{fig:visual}
\end{figure*}

\begin{figure}
    \centering
    \includegraphics[width=\linewidth]{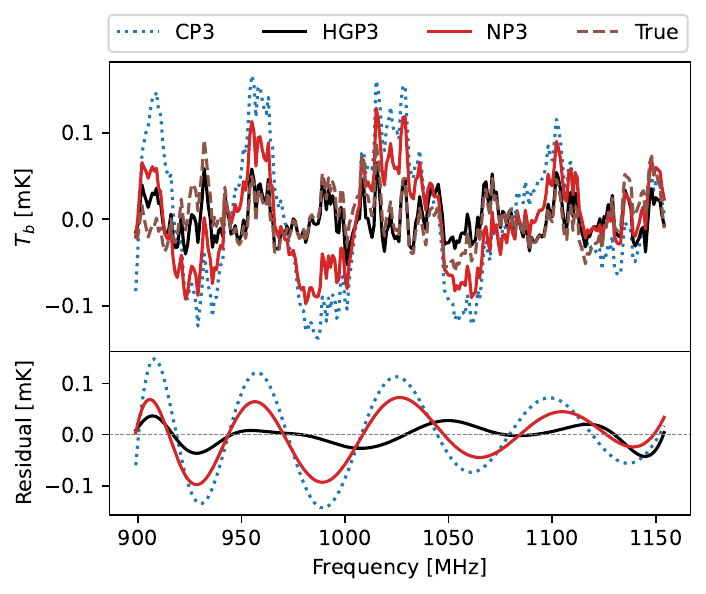}
    \caption{A comparison between the CP3, NP3 and HGP3 models of the recovered {$\bi{f}_{\rm res}$} spectrum along a single LoS {with physical coordinates $(x_{\rm pix},y_{\rm pix}) = (1374, 1408)\: {\rm Mpc}$ in the upper right corner of slices shown in Figure \ref{fig:visual}}, where the CP3 and NP3 models show significant under-cleaning of the foreground emission. The top panel shows the recovered $\bi{f}_{\rm res}$ spectrum for each of the models, plotted alongside the true $\bi{f}_{\rm res}$ spectrum, shown as a brown dashed line. The bottom panel shows the corresponding residuals with respect to the truth for each model.}
    \label{fig:res}
\end{figure}

\begin{figure}
    \centering
    \includegraphics[width=\linewidth]{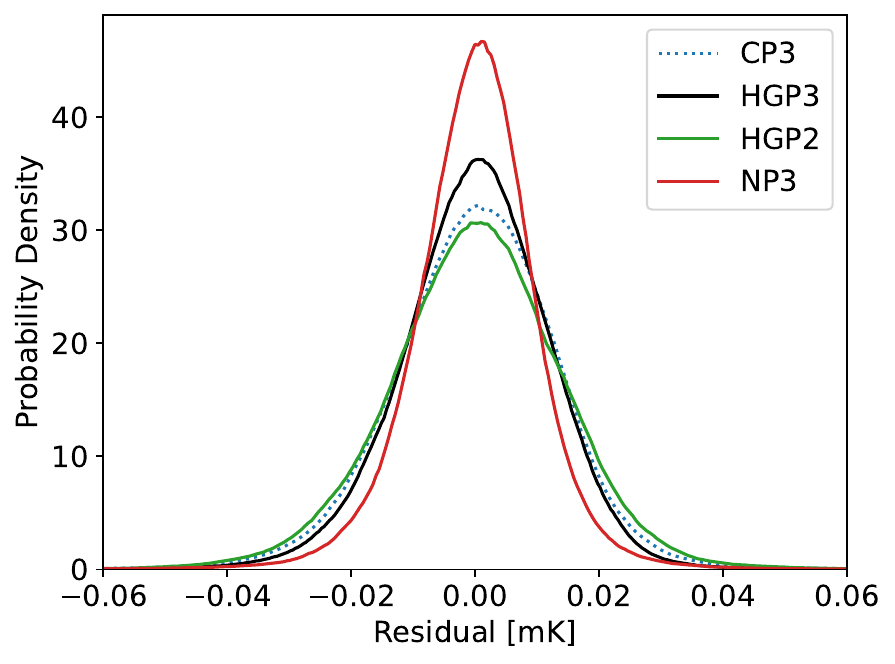}
    \caption{The probability density functions for the residuals, {$\bi{r} = {\bi{f}_{\rm res} - \bi{f}_{\rm True}}$}, of the recovered {$\bi{f}_{\rm res}$} signal cubes for the different GP models. We show the residual distribution for the CP3 (blue dotted line), HGP3 (black solid line), HGP2 (green solid line), and NP3 (red solid line), respectively. 
    The HGP3 and NP3 residual distributions are narrower than the CP3 residual distribution. However, this global distribution does not fully highlight the minority of regions where the CP3 and NP3 models result in significant under-cleaning.}
    \label{fig:compare}
\end{figure}

\begin{figure}
    \centering
    \includegraphics[width=\linewidth]{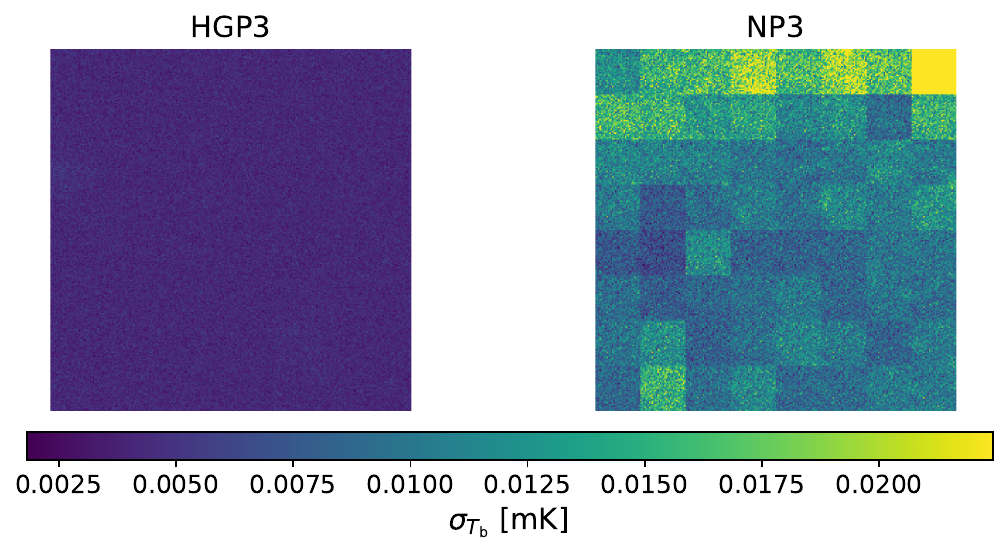}
    \caption{The $1\sigma$ uncertainty on the recovered {$\bi{f}_{\rm res}$} signal with the HGP3 (left) and NP3 models (right) in each pixel at 1019 MHz. Both plots are shown on the same color scale. The NP3 model uncertainty is about ten times greater than the HGP3 uncertainty. This is expected behavior when comparing hierarchical and no pooling approaches, with the lack of hyper-prior regularization resulting in a noisier signal recovery for the NP3 model.}
    \label{fig:uncertainty}
\end{figure}

\section{Results}\label{sec:results}

In this section, we show the {$\bi{f}_{\rm res}$} signal recovery results, using the various GP models discussed in Section \ref{sec:gpmodels}. In Section \ref{subsec: HI cubes} we discuss the recovery of the {$\bi{f}_{\rm res}$} signal cubes in the pixel domain, {evaluated by analyzing the residuals $\bi{r} = {\bi{f}_{\rm res} - \bi{f}_{\rm True}}$, where $\bi{f}_{\rm True}$ is the true $\bi{f}_{\rm res}$ signal}, alongside the predictive performance of the models in this domain. In Section \ref{subsec: summary stats} we compare the recovery of the power spectrum and scattering transform coefficient summary statistics.

\subsection{Recovered {$\bi{f}_{\rm res}$} Image Cubes}\label{subsec: HI cubes}
\begin{table}
\centering
\setlength{\tabcolsep}{4mm}{
\begin{tabular}{|l|c|}
\hline
\textbf{GP model} & \textbf{Normalized result} \\
\hline
CP3 & 1.0000 \\
HGP3 & 1.0027 \\
HGP2 & 1.0015 \\
NP3\tablenotemark{*} & 1.0146 \\
\hline
\end{tabular}}
\tablenotetext{*}{Only 1\% of the Pareto-$\hat{k}$ diagnostics are good ($\hat{k}\leq 0.5$), suggesting this value is unreliable and the NP model is misspecified.}
\caption{Leave-One-Out Cross-Validation (LOO-CV) Results: This table presents the ratio of estimated the log pointwise predictive density ($\mathrm{elpd}_\mathrm{loo}$) between different GP models and CP3 results. These results provide insights into each model's predictive performance, with higher values indicating better predictive performance. The standard error on the $\mathrm{elpd}_\mathrm{loo}$ estimates was $\sim 1\%$ for all the models.}

\label{table:elpd}

\end{table}

We first consider our benchmark CP models. In the second column of Figure \ref{fig:visual} we show the recovered {$\bi{f}_{\rm res}$} brightness temperature field at a frequency of $\nu=1019\,\rm{MHz}$. For the CP3 model, we can see good recovery over much of the observed field. However, there is a region of visible under-cleaning in the upper right corner of the field. This is further illustrated in Figure \ref{fig:res}, where we show the recovered {$\bi{f}_{\rm res}$} signal along a given LoS. In this case, it can be seen that the CP3 model results in a significant excess signal, caused by residual foregrounds after cleaning. We also show the recovered {$\bi{f}_{\rm res}$} field for the CP2 model {in Figure \ref{fig:visual}}. In this case, we see poor recovery of much of the sky, which is consistent with previous analyses in \citetalias{Soares_2022}. {We discuss the performance of the CP2 model further in Appendix \ref{sec: cp2 performance}, and given the significantly worse signal recovery, do not consider it further in the main text.}

The recovered {$\bi{f}_{\rm res}$} brightness temperature fields are shown for the NP and HGP models in the third and fourth columns of Figure \ref{fig:visual} respectively. For the NP3 and HGP3 models we again obtain good recovery over much of the sky, albeit with superpixel boundary artifacts due to the division of the data cube. This is particularly apparent for the NP3 model, where the \HI{} GP kernel had to be divided over the data subsets used for inference. Whilst we still obtain improved residuals and recovery of summary statistics compared to the CP3 model, these artifacts could be reduced or removed by considering memory-efficient GP approximations that would allow us to avoid the need to use data subsets or superpixels \citep{solin2020hilbert,riutort2023practical}. We defer the analysis of potential GP approximations to future work. The visible under-cleaning seen in the upper right of the {$\bi{f}_{\rm res}$} field for the CP3 model is still apparent with the NP3 model. In contrast, the HGP3 model eliminates this region of under-cleaning. This can be seen clearly in Figure \ref{fig:res}, where the CP3 and NP3 signals are significantly undercleaned for the relevant LoS, whilst the HGP3 signal much more closely matches the true {$\bi{f}_{\rm res}$} signal from the simulations.

Whilst both the NP3 and HGP3 models allow for spatial variation in foreground kernel parameters, the NP3 model lacks any regularization from the global hyper-prior in the HGP3 model, and does not share information between different LoS when fitting the foreground signal. The extra freedom of the NP model and lack of regularization can result in overfitting in reaction to noisy and/or anomalous data.  

The HGP2 model is able to recover the {$\bi{f}_{\rm res}$} signal better than the CP2 model, although with a degraded performance compared to the HGP3 model. The HGP2 model has more freedom than the CP2 model, by allowing foreground kernel parameters to vary spatially, whilst regularizing that variation through the global hyper-prior, resulting in a model that is more robust to misspecification in this instance.  We were unable to reach convergence with the NP2 model after extensive runs with the NUTS algorithm. This is not unexpected, given that the NP model does not share information across all LoS, which can result in convergence challenges for the misspecified NP2 model. We therefore do not consider the NP2 results further in this work.

In Figure \ref{fig:compare} we show the pixel-level residual distributions for the CP3 model, the NP3 model, and the HGP2 and HGP3 models. We find that the NP3 model has the narrowest residual distribution, followed by the HGP3 model, the CP3 model, and the HGP2 model respectively. However, the total residual distribution alone is not wholly indicative of the model performance. Indeed, it was clear that there were regions of the sky where both the NP3 and CP3 models struggled to recover the correct {$\bi{f}_{\rm res}$} signal, whilst the HGP models could. These features are hidden in the residual distribution for all pixels in the data set. We also see that the HGP2 residual distribution is wider than that of the CP3 model, demonstrating the importance of including all relevant kernel contributions in foreground modeling. In Figure \ref{fig:uncertainty} we also show the recovered $1\sigma$ uncertainty on the recovered {$\bi{f}_{\rm res}$} signal for the HGP3 and NP3 models at a given frequency slice, estimated by evaluating the ensemble variance of the posterior predictive {$\bi{f}_{\rm res}$} signal samples. The typical uncertainty on the recovered NP3 signal is about ten times larger than the uncertainty on the signal recovered by the HGP3 model. This is to be expected, given that the NP3 model has a larger number of free parameters that are not regularized through a global hyper-prior, resulting in a noisier signal recovery. Indeed, such behavior is typical of hierarchical models compared to no-pooling models \citep{gelman2006data}.

To make a more rigorous comparison of the model performance, we consider the leave-one-out cross-validation (LOO-CV) statistics \citep{vehtari2017practical}. LOO-CV is a widely used method in statistical modeling and machine learning for evaluating the predictive performance of models. In addition to model performance, LOO-CV can also provide insights into the stability and robustness of a model. Given some data set $y=\{y_1,\ldots,y_N\}$, and a model with parameters $\theta$, LOO-CV seeks to estimate the leave-one-out log predictive density of the model,
\begin{equation}
    \mathrm{elpd}_{\mathrm{loo}} = \sum_{i=1}^N \log p(y_i| y_{-i}),
    \label{eqn: elpd loo}
\end{equation}
where $y_{-i}$ is the data set left after removing the datapoint $y_i$. The pointwise predictive density is given by
\begin{equation}
    p(y_i|y_{-i}) = \int p(y_i|\theta)p(\theta|y_{-i})\mathrm{d}\theta.
\end{equation}
An estimator for $\rm{elpd}_{\rm loo}$ can be constructed using posterior samples and Pareto smoothed importance sampling \citep{2015arXiv150702646V}. In essence, $\rm{elpd}_{\rm loo}$ evaluates the performance of the model in predicting held out data.

In Table \ref{table:elpd} we show the $\rm{elpd}_{\rm loo}$ values for the CP3, NP3, HGP3 and HGP2 models, all normalized by the $\rm{elpd}_{\rm loo}$ value for the CP3 model. The NP3, HGP3 and HGP2 models all have higher $\rm{elpd}_{\rm loo}$ values than the CP3 model. Whilst the NP3 model has a higher value for $\rm{elpd}_{\rm loo}$ than the HGP3 and HGP2 models, this estimate should be treated with caution. Only $\sim 1\%$ of the Pareto-$\hat{k}$ statistics are less than 0.5 for the NP3 model. The Pareto-$\hat{k}$ statistic can be used as a diagnostic for the variance of the importance sampling estimator used to evaluate $\rm{elpd}_{\rm loo}$, with values of $\hat{k}<0.5$ being necessary to ensure finite variance of the importance ratios and that the central limit theorem holds for estimating $\rm{elpd}_{\rm loo}$ \citep{2015arXiv150702646V}. Therefore, the $\rm{elpd}_{\rm loo}$ values for the NP model should be viewed as unreliable. This behavior commonly manifests as a result of strong model misspecification. This could be a result of assumptions such as the total independence of foreground emission along every LoS in the NP model. In reality, we know foreground emission will have spatial correlations. Furthermore, the division of the full data set into subsets due to the memory limitations means that the \HI{} kernel was treated independently for each data subset, which does not conform to our expectations for the cosmological signal.

\subsection{Summary Statistics Recovery}\label{subsec: summary stats}

In this section, we consider the recovery of summary statistics from the foreground-cleaned {$\bi{f}_{\rm res}$} maps. In Section \ref{subsec: power spectra} we study the power spectrum (PS) recovery, and in Section \ref{sec:st} we examine the recovery of scattering transform coefficients.

\subsubsection{The 21 cm Power Spectrum}\label{subsec: power spectra}
\begin{figure*}
    \centering
    \includegraphics[width=\linewidth]{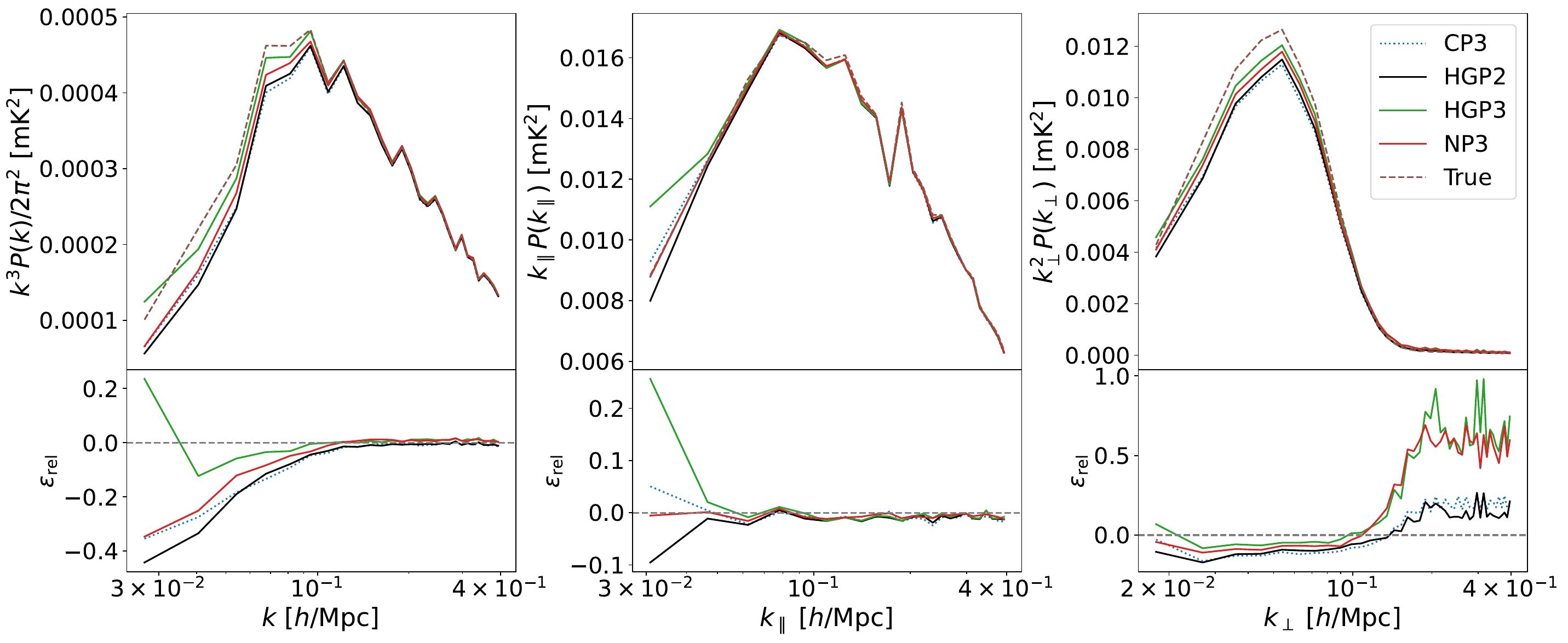}
    \caption{\textbf{Left}: The upper panel shows the recovered spherically averaged power spectra for the different GP models at $0.23<z<0.58$, and the relative error compared to the true PS is shown in the lower panel. We show the true {$\bi{f}_{\rm res}$} power spectrum (brown dashed line), the result for the CP3 model (blue dotted line), the HGP3 model (black solid line), the HGP2 model (green solid line), and the NP3 model (red solid line), respectively.  \textbf{Middle}: Recovered radial power spectra and relative error. \textbf{Right}: Recovered transverse power spectra and relative error. All of these power spectra are shown here without bias correction.}
    \label{fig:psdefault}
\end{figure*}

The 21~cm PS is perhaps the most well-studied summary statistic, capturing rich cosmological and astrophysical information \citep[e.g.,][]{2004ApJ...608..622Z,2017MNRAS.472.2651G}. {The $n$-dimensional power spectrum is defined through
\begin{equation}
    \langle\Tilde{\bi{f}}_{\rm res}(\mathbf{k}), \Tilde{\bi{f}}_{\rm res}(\mathbf{k'})\rangle = (2\pi)^{n}\delta^{(n)}(\mathbf{k}+\mathbf{k'})P(k=|\mathbf{k}|),
\end{equation}
where $\Tilde{\bi{f}}_{\rm res}(\mathbf{k})$ is the Fourier transform of an $n$-dimensional residual field $\bi{f}_{\rm res}$, consisting of cosmological 21 cm signal and instrumental noise, $P(k)$ is the corresponding PS for some wavenumber $k$, and $\delta^n(\cdot)$ is the Kronecker-delta function. We calculate three different PS: The first is the 1D radial PS, $P(k_{\parallel})$, where $k_{\parallel}$ is the corresponding radial wavenumber. In this case we evaluated the 1D PS of the recovered $\bi{f}_{\rm res}$ field along each LoS before averaging the PS for each LoS over the full data cube. The second is the 2D transverse PS, $P(k_{\perp})$, where $k_{\perp}$ is the corresponding transverse wavenumber. In this case the 2D PS was evaluated over the 2D fields corresponding to each frequency slice before averaging the PS for each frequency slice over the full data cube. The third is the 3D spherically averaged PS, $P(k)$, where $k$ is the wavenumber associated with the full 3D $\bi{f}_{\rm res}$ field. In this case the PS was evaluated over the full 3D data cube. For visualization purposes, we show the dimensionless PS i.e., $k_{\parallel}P(k_{\parallel})$, $k_{\perp}^2P(k_{\perp})$, and $k^3P(k)/2\pi^2$.}

{We choose the minimum wavenumber for the 1D radial PS to be $k_{\parallel,\rm min} = 2\pi L_z^{-1}$. For the 2D transverse PS, the minimum scale is set to $k_{\perp,\rm min} = 2\pi (L_x^2+L_y^2)^{-1/2}$ and for the 3D PS we set $k_{\rm min} = 2\pi V^{-1/3}$, where $V$ is the volume of the data cube. The bin-width was set to $2k_{*,\min}$, i.e., twice the minimum wavenumber for each case, and the maximum wavenumber was set to $0.4\: h/{\rm Mpc}$ for all three cases.}

Here we investigated the PS recovery from the foreground cleaned {$\bi{f}_{\rm res}$} maps, showing the recovered PS in Figure \ref{fig:psdefault}. For the spherically averaged PS, the smallest relative error is obtained with the HGP2 model. The NP3 model achieves better recovery than the HGP3 model, which obtains results comparable to the CP3 model.

For the 1D PS along LoS, all four models attain comparable recovery on small scales, with differences only being apparent on large scales. The NP3 model achieves the lowest relative error in this case, and the CP3 and HGP3 models have comparable relative errors on large scales. The HGP2 model significantly overestimates the PS on large scales, indicating significant uncleaned foreground components in the recovered {$\bi{f}_{\rm res}$} signal. Using one foreground kernel often fails to describe the polarization leakage effects, which have a different frequency behavior compared to the smooth astrophysical foregrounds in total intensity.

For the 2D transverse PS, all methods underestimated the power on scales $k_\perp\lesssim 0.1h/\rm Mpc$, indicating mismodeling of foreground spatial variations. The HGP2 model gives the best recovery on the scales $k_\perp\lesssim 0.1h/\rm Mpc$, indicating that it was better able to preserve Gaussian spatial fluctuations on these scales. The HGP3 and CP3 models attain comparable relative error in this case, with the NP3 model attaining a smaller relative error. This would suggest that the regularization through the hyper-prior with the HGP3 model, and the complete pooling of all foreground kernel parameters for the CP3 model, act to reduce spatial fluctuations compared to the NP3 model. On smaller scales, the HGP3 and CP3 models attain smaller relative errors compared to the HGP2 and NP3 models. However, on these small scales, the instrumental beam suppresses the PS close to zero, with the limitations of floating point accuracy making comparisons through the relative error unreliable.

Whilst the PS recovery seen here would indicate good performance for the HGP2 and NP3 models, this should be treated with some caution, given that these PS estimates have not undergone any bias correction. This issue is considered in more detail in Section \ref{subsec:bias}. However, it is clear from the recovered 2D transverse PS that all methods fail to properly describe spatial variations in the foreground signal. Whilst the NP3 model allows for greater freedom, and hence spatial variations, this method still mismodels the true foreground emission by assuming total independence between every LoS, as discussed previously in the context of the LOO-CV results. Whilst the HGP3 model allows for spatial variations in the foreground emission kernels, it is assumed that these kernel parameters are all drawn from some global hyper-prior. This is an overly simplistic model. In reality, we would expect foreground kernel parameters to display local spatial correlations. An improved spatial prior should therefore account for these local correlations. Given that modeling local spatial correlations introduces significant additional computational challenges, we leave an exploration of such models to future work.

\subsubsection{Wavelet Scattering Transform Coefficients}
\label{sec:st}
\begin{figure*}
    \centering
       \includegraphics[width=0.305\textwidth]{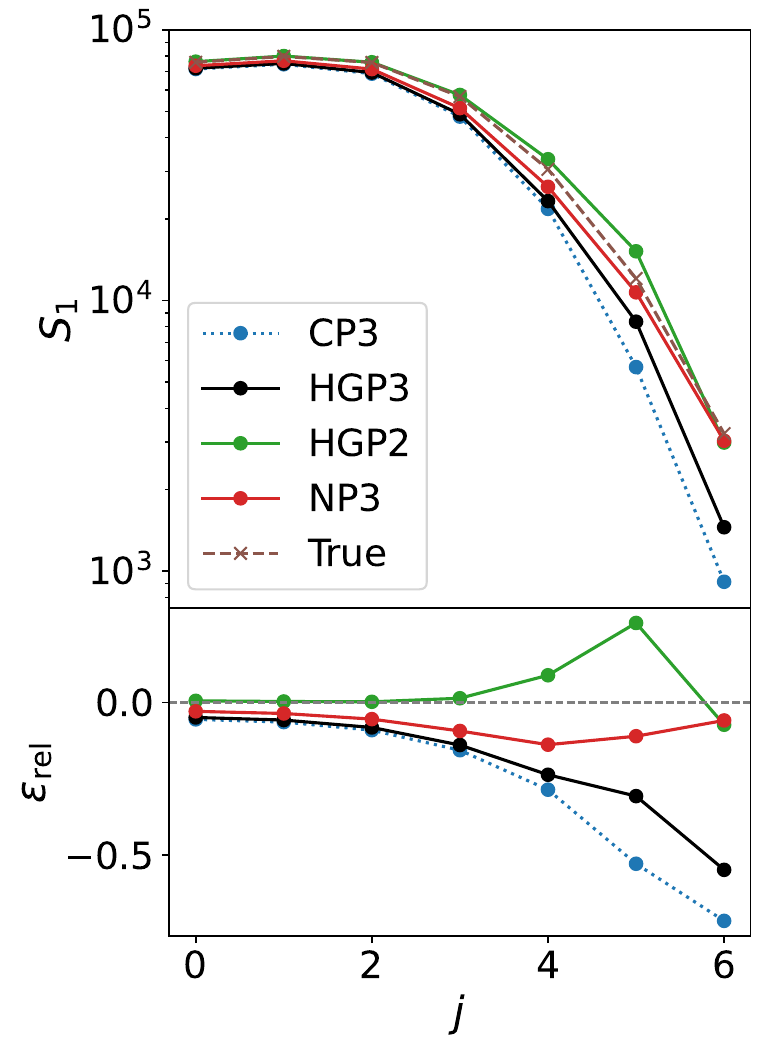}
        \includegraphics[width=0.675\textwidth]{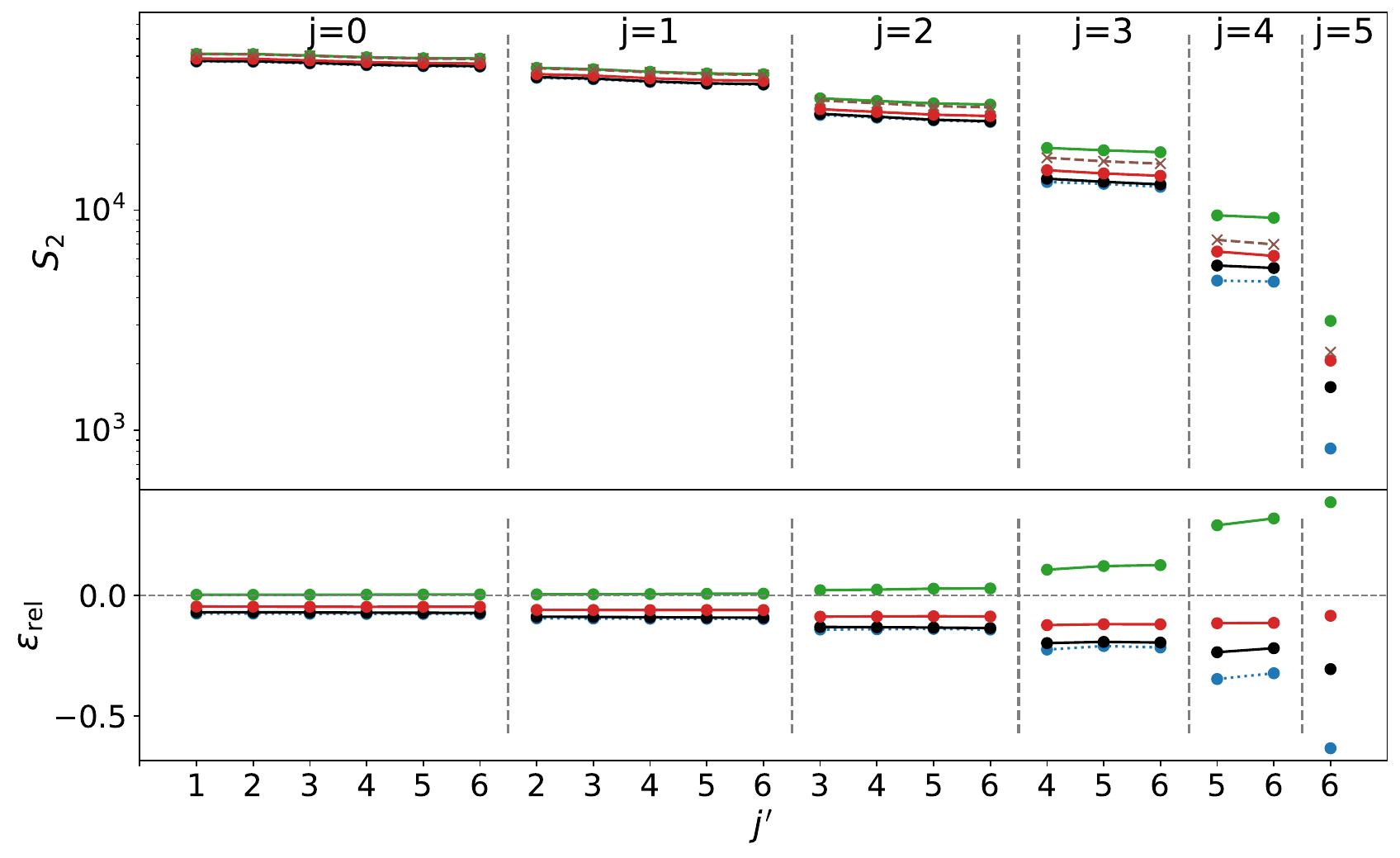}
    \caption{The first-order ($S_1$, left) and second-order ($S_2$, right) ST coefficients for the different GP models. The ST coefficients are averaged over $0 \leq \ell \leq 6$ and evaluated at $q=1$. We show the ST coefficients for the true field (brown crosses), the CP3 model (blue circles), the HGP3 model (black circles), the HGP2 model (green circles), and the NP3 model (red circles), respectively. The top panel shows the ST coefficients and the bottom panels show the relative error with respect to the ST coefficients for the true field.}
    \label{fig:st}
\end{figure*}

The PS characterizes Gaussian information in the relevant field. However, the full \HI{} field is non-Gaussian, with any non-Gaussian information beyond the PS summary. Recently, the wavelet scattering transformation (WST) has emerged as a new approach to extract information from cosmological fields \citep[e.g.][]{2020MNRAS.499.5902C, 2023MNRAS.519.5288G,2023arXiv230604689D} and is more informative than the poly-spectrum in Fourier space \citep{2023arXiv230704994S,2021arXiv211201288C}. To examine the recovery of non-Gaussian summary statistics, we calculate the three-dimensional solid harmonic WST coefficients \citep[e.g.][]{2023arXiv231017602Z,2018JChPh.148x1732E} of the recovered {$\bi{f}_{\rm res}$} signal cubes.

We briefly summarize the solid harmonic WST here. The solid harmonic WST convolves the original field $\mathbf{d}(\mathbf{x})$ with a cascade of solid harmonic wavelets defined as
\begin{equation}
\begin{aligned}
    &\psi_{\ell}^{m}(\mathbf{x})=\frac{1}{(\sqrt{2 \pi})^{3}} e^{-|x|^{2} / 2}|\mathbf{x}|^{\ell} Y_{\ell}^{m}\left(\frac{\mathbf{x}}{|\mathbf{x}|}\right),\\
    &\psi_{j, \ell}^{m}(\mathbf{x})=2^{-3 j} \psi_{\ell}^{m}\left(2^{-j} \mathbf{x}\right),
\end{aligned}
\label{eqn:wavelet}
\end{equation}
where $\mathbf{x}$ is the spatial coordinates, $Y_{\ell}^{m}$ is the three-dimensional spherical harmonic function and $\psi_{j, \ell}^{m}(\mathbf{x})$ is the convolution kernel parameterized by $j,\ell$ and $m$. Here, $j$ gives the spatial scale of the kernel as $2^{j}$, whilst $\ell$ and $m$ characterize the angular scale. Nonlinear moduli are applied to the convolved field as
\begin{equation}
U[j, \ell] \mathbf{d}(\mathbf{x})=\left(\sum_{m=-\ell}^{\ell}\left|\mathbf{d} * \psi_{j, \ell}^{m}(\mathbf{x})\right|^{2}\right)^{1 / 2}\,,
\label{eq:E-norm}
\end{equation}
where the field $\mathbf{d}(\mathbf{x})$ is convolved (denoted by ``$*$'') with $\psi_{j, \ell}^{m}(\mathbf{x})$.   
The first-order solid harmonic WST coefficients (hereafter first-order ST coefficients) are obtained by integrating the field as
\begin{equation}\label{eq:shw1}
S_1 \left[\mathbf{d} ; j, \ell, q \right] =\int_{\mathbb{R}^{3}}|U[j, \ell] \mathbf{d}(\mathbf{x})|^{q} \mathrm{d}^3\mathbf{x}.
\end{equation}
In order to capture information across multiple scales, the first-order solid harmonic WST field $U[j, \ell] \mathbf{d}(\mathbf{x})$ is convolved again with a kernel with different $j^\prime >j$ but with the same $\ell$. The second-order ST coefficients are then given by
\begin{equation}
    S_2 \left[\mathbf{d} ; j,j^\prime, \ell, q \right] =\int_{\mathbb{R}^{3}}\left|U[j^\prime, \ell] U[j, \ell] \mathbf{d}(\mathbf{x})\right|^{q} \mathrm{d}^3\mathbf{x},\quad j<j^\prime.
\end{equation}

In our analysis, we choose $j_{\max}=6$ to explore features on small to large scales, select $\ell_{\max}=6$ to explore different angular directions, and set $q=1$ in order to characterize the convolved and modulated field without any amplification. The ST coefficients are calculated using \texttt{Kymatio}\footnote{\url{https://www.kymat.io/}} \citep{2018arXiv181211214A}. The ST coefficients are averaged over the angular scales $\ell$, as the averaged ST coefficients are informative \citep{2023arXiv231017602Z}.

The recovered ST coefficients obtained with the different GP models are shown in Figure \ref{fig:st}. For the first-order ST coefficients $S_1$, the HGP2 model obtains the smallest relative error on small scales when $j\leq 4$. The relative errors obtained with the NP3 and HGP3 models are worse than the HGP2 model, albeit better than those obtained with the CP3 model. For $j=5$, the NP3 model achieves the smallest relative error, with the CP3 model having the worst relative error. For the largest scale $j=6$, the HGP2 and NP3 models have almost identical recovery, attaining the smallest relative error, with the HGP3 model still having a smaller relative error than the CP3 model. 

For the second-order ST coefficients $S_2$, the results follow a similar pattern. For small-scale first-order maps with $j\leq 2$, the HGP2 first-order maps are almost identical to the true-field first-order map. For $j\geq 4$ the NP3 model has the smallest relative error, with the worst relative error obtained by the CP3 model for all scales.

As with the PS estimates presented in Section \ref{subsec: power spectra}, the ST coefficients show good results for the HGP2 model and NP3 model, with the HGP3 model also having better recovery of non-Gaussian summary statistics than the CP3 model. In making complete comparisons of the model performance in recovering the full non-Gaussian field, the LOO-CV summaries in Section \ref{subsec: HI cubes} should be considered with the discussion therein. However, the recovered ST coefficients do provide a useful check on the non-Gaussian information recovery. As discussed with the PS recovery, the performance of the HGP3 model would be improved by using a more realistic spatial prior, with the global hyperprior used in this work being an incomplete model of foreground spatial correlations.

\section{Discussions}\label{sec:dis}

In this section, we consider issues related to the cosmological signal recovery along with modeling and computational choices. In Section \ref{subsec:bias}, we study the effect of bias corrections on power spectrum recovery. In Section \ref{subsec:super-pixel}, we discuss the choice of superpixel size used for the HGP2 and HGP3 models. In Section \ref{subsec:map vs HMC}, we compare the {$\bi{f}_{\rm res}$} signal recovery when using MAP-based inference of the kernel hyperparameters compared to the fully Bayesian approach we use throughout this work. {In Section \ref{sec: bad pixel}, we discuss where the NP3 and CP3 models have poor signal recovery, and why these models can be prone to such degraded signal recovery.}

\subsection{Bias Correction}\label{subsec:bias}
\begin{figure*}
    \centering
       \includegraphics[width=0.95\textwidth]{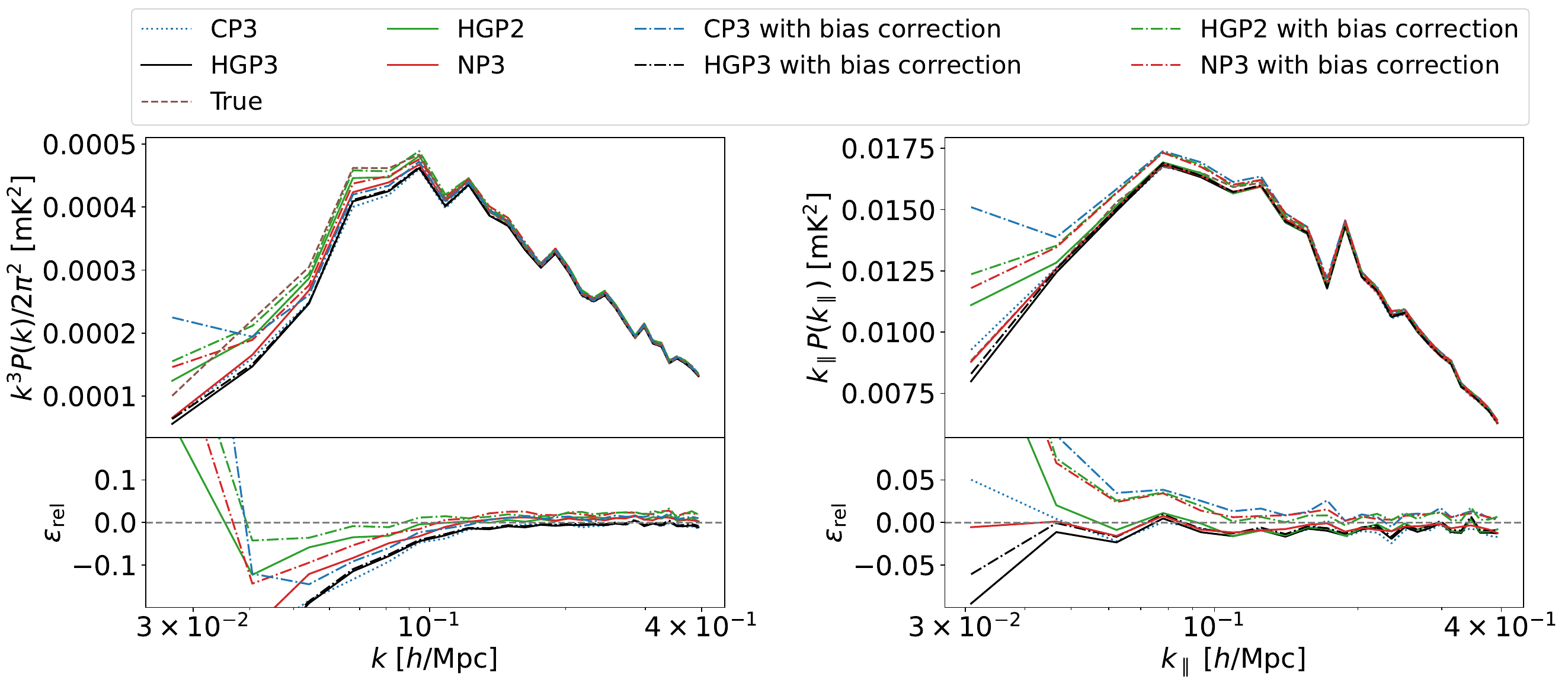}
    \caption{Spherically averaged PS (left) and radial PS (right) with and without bias correction for different GP models. The uncorrected and true PS are shown with the same color type and line type as in Figure \ref{fig:psdefault}, while the bias-corrected PS are shown with the dash-dotted lines and with the same color type as the uncorrected results for each model. Bias correction results in overestimation of the spherically averaged PS on large scales for the CP3, HGP2 and NP3 models. The bias-corrected radial PS is overestimated on all scales for the CP3, HGP2 and NP3 models.}
    \label{fig:psbias}
\end{figure*}
As discussed in Section \ref{sec: gp models}, the 21~cm signal is recovered by subtracting $\mathbb{E}[f_{\rm fg}]$ from the total observed signal, whilst the foreground covariance $\mathrm{cov}[f_{\rm{fg}}]$ is ignored. However, \citet{2020MNRAS.493.1662M} shows that this can result in a bias in the recovery as
\begin{equation}
\begin{aligned}
    \left< f_{\rm 21cm,\, est}f_{\rm 21cm,\, est}^T\right> &= \\
\left< f_{\rm 21cm,\, est}\right.&\left.f_{\rm 21cm,\, est}^T\right>_{\rm unbiasd} - {\rm cov}[f_{\rm fg}].
\end{aligned}
\end{equation}
\citet{2020MNRAS.493.1662M} shows that to compensate, an additive bias correction for the radial PS can be performed with the Monte Carlo procedure as follows.
\begin{itemize}
    \item[1.] Generate several random samples from a multivariate Gaussian distribution, with zero mean and a covariance given by ${\rm cov}[f_{\rm fg}]$. Each sample has the same size as our data cube.
    \item[2.] Calculate and average the radial PS over these realizations.
    \item[3.] Add the averaged PS to the residual PS obtained with GP foreground subtraction. 
\end{itemize} 
The procedure for the spherically averaged PS is similar, except that in the second step the calculated radial PS is binned into the $k$-bins. This ignores the $k_\perp$ modes when estimating the bias for the spherically averaged PS. For the modeling considered in this paper, the covariance matrix $\mathrm{cov}[f_{\rm{fg}}]$ does not contain information regarding the foreground signal covariance in the transverse direction \citepalias{Soares_2022}.

In this work, we adopt the method described in \citet{2020MNRAS.493.1662M} to examine the effects of bias correction. However, \citet{2021MNRAS.501.1463K} highlighted the potential failure of this correction in accurately retrieving the true 21~cm PS from the EoR due to misestimation of the EoR signal covariance, and instead proposed a multiplicative correction that cannot under-predict the EoR power spectrum, but this approach requires perfect knowledge of the foreground covariance. However, in this work our methods are based on the idealized simulations used in \citetalias{Soares_2022}, which justifies our estimation of the data covariance. We defer a detailed exploration of the corrections proposed by \citet{2021MNRAS.501.1463K} to future work.

For the HGP and NP models, we adopted multiple covariance matrices to model the foreground emission. We should therefore calculate the bias correction for superpixels along each LoS (for HGP) or for pixels (for NP) independently. However, in practice small values of ${\rm cov}[f_{\rm fg}]$ require better computational precision than double precision. This significantly slows the calculation down, making a full bias correction for the NP model intractable with our available computational resources. For simplicity, we calculate a global ${\rm cov}[f_{\rm fg}]$ with averaged hyperparameters as an approximation for the NP3 model and leave the analysis of full corrections to future work.

The results of the PS bias correction are shown in Figure \ref{fig:psbias}. 
On scales $k\gtrsim 0.05\,h/\rm Mpc$, the spherically averaged PS for all models are in agreement, and the effect of bias correction is negligible. However, the bias correction results in significant overestimation on larger scales. This correction is most prominent for the CP3 model and least significant for the HGP3 model. The impact of bias correction can be seen more directly in the radial PS, where bias correction results in overestimation of the radial PS with the HGP2, NP3, and CP3 models on all scales. This is not the case for the HGP3 model, where the bias correction is again small, and we do not observe an overestimation of the radial PS on large scales. The overcorrection through the $k_\parallel$ modes applied to the HGP2, NP3, and CP3 models manifests itself as an apparently improved recovery of the spherically averaged PS on intermediate scales. However, in this case, the over-correction applied to the $k_\parallel$ modes acts against the poor recovery of the $k_\perp$ modes for all methods on these scales. As discussed previously, further improvements in PS recovery, particularly for transverse PS, can be made by considering more physically motivated spatial priors. 

\begin{figure}
    \centering
    \includegraphics[width=\linewidth]{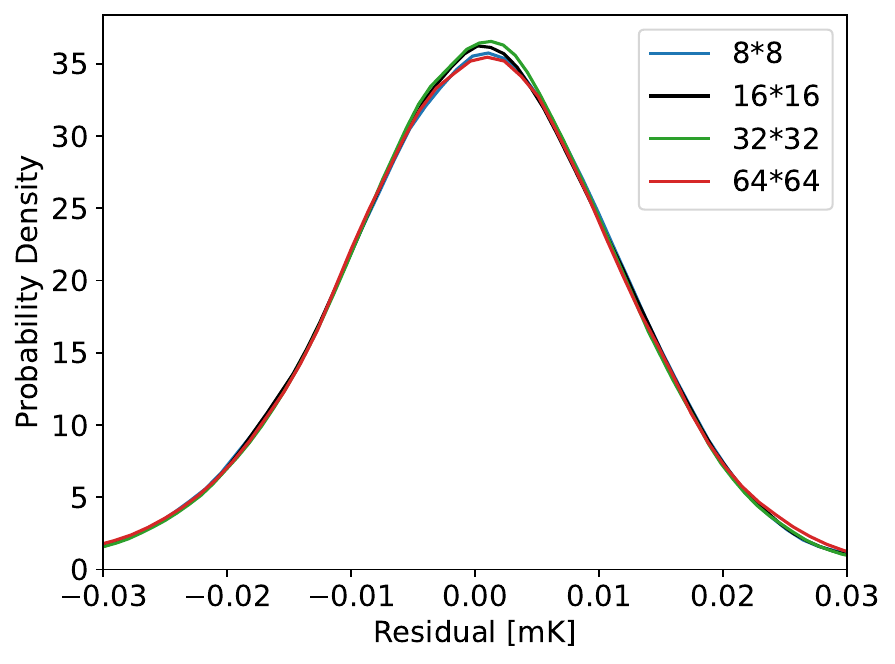}
    \caption{Residual distributions of the recovered {$\bi{f}_{\rm res}$} signal cubes for the HGP3 model with different sizes of a single superpixel, as indicated in the legend. In the test, for a single superpixel that contains from $8\times 8$ normal pixels to $64\times 64$ normal pixels, the residual distributions are broadly the same.}
    \label{fig:suppix}
\end{figure}
\begin{figure}
    \centering
    \includegraphics[width=\linewidth]{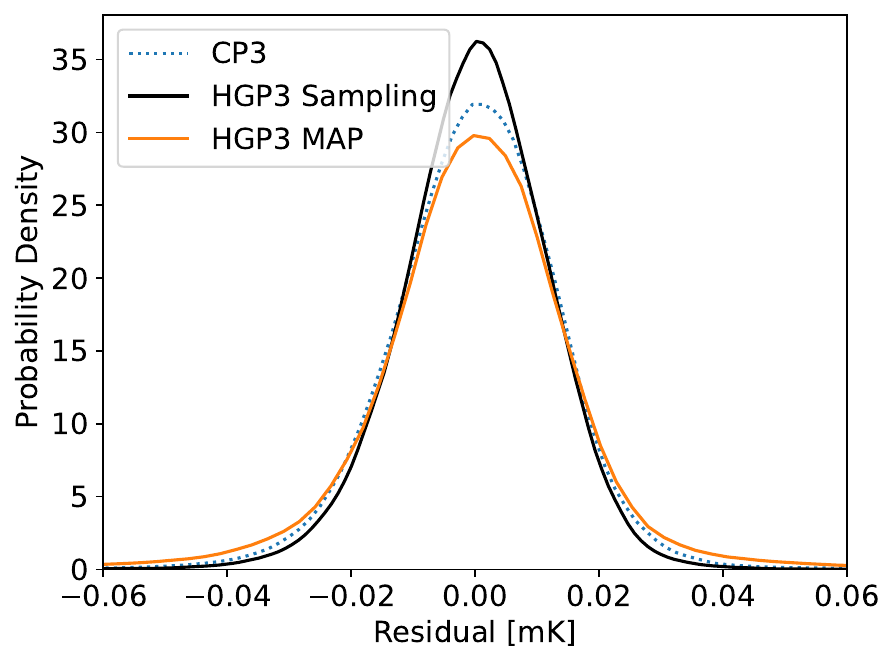}
    \caption{Residual distributions of the recovered {$\bi{f}_{\rm res}$} signal cubes for the HGP3 model using MAP inference (orange solid line) and full posterior sampling (black solid line), and, for comparison, for the CP3 model (blue dotted line),  respectively. }
    \label{fig:resmap}
\end{figure}

\subsection{Choice of Superpixel Size}\label{subsec:super-pixel}

To study the impact of choosing different sizes of a single superpixel for the HGP models, we ran the HGP3 with a single superpixel that contains from $8\times 8$ normal pixels to $64\times 64$ normal pixels. The configuration of a superpixel with $8\times 8$ normal pixels corresponds to the minimum superpixel size that fits our memory constraint because a smaller superpixel size means a larger number of superpixels, each of which has a set of free GP kernel parameters. In Figure \ref{fig:suppix} we show the residual distributions obtained with various superpixel sizes. The residual distributions are broadly the same for all superpixel sizes we test here. In this work, we therefore select a superpixel size of $16\times 16$ normal pixels, a trade-off between the computational demands of the model and allowing the model to account for more detailed spatial variations in the foreground emission. In future work, it will be interesting to consider GP approximations that would allow us to remove the need for superpixels, and hence remove any superpixel boundary effects \citep{solin2020hilbert, riutort2023practical}.

\subsection{Maximum-A-Posteriori inference vs. sampling}\label{subsec:map vs HMC}
\label{subsec:MAP vs sampling}

Throughout this work we used a fully Bayesian approach, sampling the GP posterior and calculating posterior expectations. Whilst this is computationally challenging, it has been found to result in improved inference in GP models, and in Bayesian models more generally \citep{mackay2003information, betancourt2017conceptual, lalchand2020approximate}. To compare the impact of Maximum-A-Posteriori (MAP) inference with full posterior sampling in our case, we consider performing MAP optimization of the GP kernel parameters for the HGP3 model, and compare the recovered {$\bi{f}_{\rm res}$} signal cubes with those obtained through sampling. The residual distributions for the two approaches are shown in Figure \ref{fig:resmap}, alongside the residual distribution for the CP3 model obtained with full sampling. We find that using MAP optimization to obtain a point estimate of the GP kernel parameters results in a residual distribution that is even broader than the CP3 residual distribution. This is consistent with the known properties of the MAP estimators. For the complex, high-dimensional geometry of the HGP3 model, we expect that the MAP estimate of the kernel parameters is far from the typical set, where the posterior probability mass is concentrated \citep{betancourt2017conceptual}.

\subsection{Examining Poor Signal Recovery}
\label{sec: bad pixel}
\begin{figure}
    \centering
    \includegraphics[width=\linewidth]{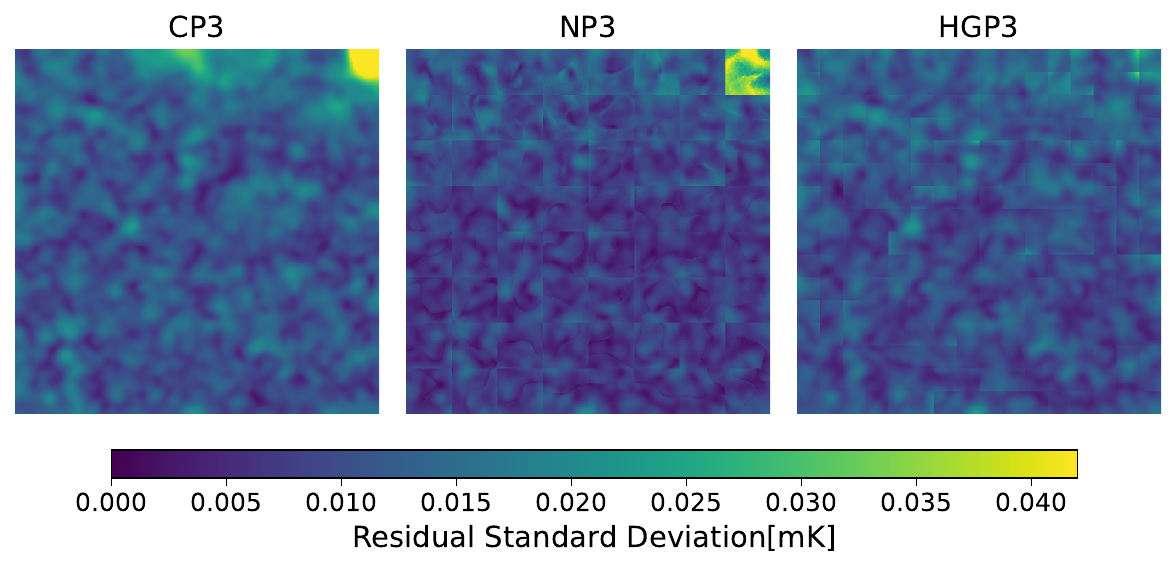}
    \caption{{The standard deviation of residuals between the recovered, foreground cleaned signal and the true signal along each LoS for the HPG3, NP3 and CP3 models. Whilst the NP3 model has lower typical residuals over much of the sky, both the NP3 and CP3 models have very large residuals in the upper right corner of the field.}}
    \label{fig:maxres}
\end{figure}

{In Figure \ref{fig:res} we showed the recovered {$\bi{f}_{\rm res}$} spectrum along a single LoS where the typical residuals obtained with the CP3 and NP3 models were much greater than those obtained with the HGP3 model. In this section we consider the residual distribution for the three models over all the LoS. In Figure \ref{fig:maxres} we show the standard deviation of residuals along the LoS for each pixel with the HGP3, NP3 and CP3 models. Consistent with the residual distributions shown in Figure \ref{fig:compare}, over most of the sky the NP3 model recovers lower residuals. However, this is not fully reflective of the model performance. In the upper right corner of the field we see a region of very poor recovery for the CP3 and NP3 models. The foreground amplitude and spectral behavior in this region are comparable to the remainder of the sky. As such, whilst it is not clear what the exact cause of the poor recovery in this region is, both the CP3 and NP3 models have properties that mean this is not necessarily surprising. The NP3 model lacks any regularization over the GP kernel parameters for each LoS. This can manifest in over-fitting and overreaction to noise, which is typical of no-pooling models in general. At the other extreme, the CP3 model applies an extreme form of regularization by assuming the GP kernel parameters are the same for every LoS, which can result in poor recovery in regions where these values are a bad fit. This behavior is also consistent with the LOO-CV results shown in Table \ref{table:elpd}. The CP3 model was found to have worse predictive performance on held-out data compared to the HGP3 model, whereas for the NP3 model the importance sampling diagnostics were indicative of model misspecification. We can also see evidence of this tendency for mismodeling or over-fitting with the NP3 and CP3 models in the number of pixels with outlier residuals. Applying a threshold of 0.042 mK, which corresponds approximately with the $3\sigma$ level of the CP3 residual distribution, we find 97 NP3 pixels and 349 CP3 pixels that exceed this threshold, whereas no HGP3 pixels do so.}

\section{Conclusions}\label{sec: conclusions}

In this work, we consider the impact of allowing for spatial variations in GP kernel parameters when modeling foreground emission in 21~cm component separation analyses. For this purpose, we consider a range of natural modeling variations that allow for different levels of spatial variation. As our baseline, we consider the CP approach adopted in previous analyses, where the foreground GP kernel parameters are assumed to be identical for every LoS. At the next level of spatial variation, we consider the HGP models, where foreground kernel parameters were allowed to vary over a set of superpixels, each with a size of $16\times 16$ normal pixels, with the kernel parameters in turn being regularized through a global hyper-prior. Finally, we consider the NP models, where foreground kernel parameters are assumed to be completely independent along every LoS, with no regularization from any hyper-prior. In all cases, we consider models with two GP kernels, abbreviated as CP2, HGP2 and NP2, and models with three GP kernels, abbreviated as CP3, HGP3 and NP3, respectively. The two-kernel models assume one GP kernel for the \HI{} emission and one GP kernel for the foreground emission. The three-kernel models have an additional foreground kernel to model polarization leakage.

We test the performance of these models against the simulated MeerKLASS-like observations, with a focus on the recovery of the {$\bi{f}_{\rm res}$ (i.e., \HI{} signal plus instrumental noise)} signal cubes. At the level of pixels, we find that both HGP3 and NP3 models have smaller residual distributions than the CP3 model, e.g.\ the standard deviation of the NP3 residual distribution is approximately $30\%$ smaller than for the CP3 residuals. The CP2 model fails to accurately recover the {$\bi{f}_{\rm res}$} signal over the sky, and the NP2 model fails to converge. In these cases, the misspecification of the two-kernel models leads to failures in foreground modeling. The HGP2 model can converge, with a comparable residual distribution to the CP3 model. In this case, the hierarchical model is robust to model misspecification.

Whilst the NP3 model has the sharpest residual distribution, there are still areas of the sky where the foreground spectrum is incorrectly modeled, similar to the CP3 model. These areas of poor signal recovery are eliminated by the HGP3 model. A more rigorous model comparison is made by performing a LOO-CV analysis and calculating the expected log predictive density of held-out data, $\mathrm{elpd}_\mathrm{loo}$. The NP3 model has the largest value of $\mathrm{elpd}_\mathrm{loo}$, followed by the HGP3, HGP2, and CP3 models. Whilst this seems to indicate that the NP3 model has the best predictive performance, the Pareto-$\hat{k}$ statistics indicate that the importance weighting procedure used to estimate $\mathrm{elpd}_\mathrm{loo}$ for the NP3 model is unreliable. This is often caused by serious model misspecification. In the case of the NP3 model, the assumption of complete independence of foreground kernel parameters along every LoS is unrealistic. Moreover, memory constraint means that the full data cube has to be divided into data subsets, with the \HI{} kernel treated independently in each subset. The lack of regularization from a hyper-prior also means that the NP3 model is prone to overfitting. The Pareto-$\hat{k}$ statistics for the remaining models are good, with the best predictive performance obtained with the HGP3 model.

In addition to the LOO-CV analysis, we also examine the recovery of summary statistics for the {$\bi{f}_{\rm res}$} field. For the PS recovery, the results for all methods agree on small scales ($k\gtrsim 0.1\,h/\rm{Mpc}$) for spherically averaged PS and radial PS, with percentage level accuracy. At large scales, the relative error increases to about $\mathcal{O}(10\%)$. 
We also investigate the effect of bias correction applied through the $k_\parallel$ modes. Its impact is apparent in the bias-corrected radial PS, where all models except the HGP3 model overestimate the radial PS on all scales. The transverse PS is underestimated by all methods on scales $k_\perp\lesssim 0.1\,h/\rm{Mpc}$, which is caused by simplified modeling of foreground kernel spatial variations. The NP3, HGP3, and HGP2 models all attain better recovery of ST coefficients compared to the CP3 model, which indicates better recovery of non-Gaussian features in the \HI{} field. 

All these results demonstrate the importance of accounting for spatial variations in foreground emission. Indeed, the HGP2 model outperforms the CP3 model in predictive performance and recovery of {$\bi{f}_{\rm res}$} field summary statistics, despite the kernel model being misspecified. However, both NP and HGP models use highly simplified models for these spatial variations, assuming either complete independence or variation through a global hyper-prior. Improving the signal recovery requires more physically motivated spatial priors that can account for local correlations in the foreground emission. This will enable improved recovery of the transverse PS and ST coefficients. However, improved spatial modeling will likely result in significant computational demands in both memory and clock time. To address this, it will be interesting to consider memory-efficient alternatives to exact GPs \citep{solin2020hilbert, riutort2023practical}, and also more efficient inference algorithms to integrate over the high-dimensional target distribution \citep{2022arXiv221208549R}. In addition to using better motivated spatial priors, performance improvements can also be obtained through improved foreground frequency covariance modeling \citep{2024MNRAS.527.3517M}. 

Alongside these modeling considerations, in future work we will test our method against more realistic mock observations. {For instance, the commonly assumed power-law synchrotron spectrum arises from the simplified assumption of a power-law energy distribution for Cosmic Ray (CR) electrons, despite observational evidence for a more complicated spectrum \citep{2019PhRvL.122j1101A}. Moreover, we expect even a true power-law spectrum to be modified by spatial averaging effects \citep{2017MNRAS.472.1195C}, and the assumption of a constant fraction of polarization leakage is a simplification. More realistic mock observations could be achieved by using synchrotron simulation codes \citep[e.g.,][]{2009A&A...495..697W, 2020ApJS..247...18W, 2024arXiv241001136D}, applied to high precision galaxy-scale simulations incorporating CRs and magnetic fields, such as those in the FIRE suite \citep{2014MNRAS.445..581H}. In addition to this, it would be worthwhile to explore the impact of more realistic instrument effects and noise modeling. All these effects will introduce additional complications for component separation, particularly regarding correlations between different LoS, meaning that effectively modeling these spatial correlations will be critical in ensuring accurate cosmological signal recovery.}

\section*{Acknowledgments}
This work is supported by the National SKA Program of China (grant No.~2020SKA0110401) and NSFC (grant No.~12250410240 and 11821303). We thank Zhaoting Chen, Meng Zhou, Ce Sui, Siyi Zhao and Yidong Xu for their useful discussions and help. We acknowledge the Tsinghua Astrophysics High-Performance Computing platform at Tsinghua University for providing computational and data storage resources that have contributed to the research results reported within this paper. 

%

\vspace{5mm}

\software{Kymatio \citep{2018arXiv181211214A}, numpyro \citep{phan2019composable,bingham2019pyro}, 
         JAX \citep{jax2018github}, 
          GPR4IM \citep{Soares_2022}, ArviZ \citep{Kumar2019}
          }




\bibliography{refs}{}

\begin{thebibliography}{}
\expandafter\ifx\csname natexlab\endcsname\relax\def\natexlab#1{#1}\fi
\providecommand{\url}[1]{\href{#1}{#1}}
\providecommand{\dodoi}[1]{doi:~\href{http://doi.org/#1}{\nolinkurl{#1}}}
\providecommand{\doeprint}[1]{\href{http://ascl.net/#1}{\nolinkurl{http://ascl.net/#1}}}
\providecommand{\doarXiv}[1]{\href{https://arxiv.org/abs/#1}{\nolinkurl{https://arxiv.org/abs/#1}}}

\bibitem[{{Aguilar} {et~al.}(2019){Aguilar}, {Ali Cavasonza}, {Alpat},
  {Ambrosi}, {Arruda}, {Attig}, {Azzarello}, {Bachlechner}, {Barao}, {Barrau},
  {Barrin}, {Bartoloni}, {Basara}, {Ba{\c{s}}e{\v{g}}mez-du Pree}, {Battiston},
  {Becker}, {Behlmann}, {Beischer}, {Berdugo}, {Bertucci}, {Bindi}, {de Boer},
  {Bollweg}, {Borgia}, {Boschini}, {Bourquin}, {Bueno}, {Burger}, {Burger},
  {Cai}, {Capell}, {Caroff}, {Casaus}, {Castellini}, {Cervelli}, {Chang},
  {Chen}, {Chen}, {Chen}, {Cheng}, {Chou}, {Choutko}, {Chung}, {Clark},
  {Coignet}, {Consolandi}, {Contin}, {Corti}, {Crispoltoni}, {Cui}, {Dadzie},
  {Dai}, {Datta}, {Delgado}, {Della Torre}, {Demirk{\"o}z}, {Derome}, {Di
  Falco}, {Di Felice}, {Dimiccoli}, {D{\'\i}az}, {von Doetinchem}, {Dong},
  {Donnini}, {Duranti}, {Egorov}, {Eline}, {Eronen}, {Feng}, {Fiandrini},
  {Fisher}, {Formato}, {Galaktionov}, {Garc{\'\i}a-L{\'o}pez}, {Gargiulo},
  {Gast}, {Gebauer}, {Gervasi}, {Giovacchini}, {G{\'o}mez-Coral}, {Gong},
  {Goy}, {Grabski}, {Grandi}, {Graziani}, {Guo}, {Haino}, {Han}, {He}, {Heil},
  {Hsieh}, {Huang}, {Huang}, {Incagli}, {Jia}, {Jinchi}, {Kanishev}, {Khiali},
  {Kirn}, {Konak}, {Kounina}, {Kounine}, {Koutsenko}, {Kulemzin}, {La Vacca},
  {Laudi}, {Laurenti}, {Lazzizzera}, {Lebedev}, {Lee}, {Lee}, {Leluc}, {Li},
  {Li}, {Li}, {Li}, {Light}, {Lin}, {Lippert}, {Liu}, {Liu}, {Liu}, {Lu}, {Lu},
  {Luebelsmeyer}, {Luo}, {Luo}, {Luo}, {Lyu}, {Machate}, {Ma{\~n}{\'a}},
  {Mar{\'\i}n}, {Martin}, {Mart{\'\i}nez}, {Masi}, {Maurin}, {Menchaca-Rocha},
  {Meng}, {Mo}, {Molero}, {Mott}, {Mussolin}, {Nelson}, {Ni}, {Nikonov},
  {Nozzoli}, {Oliva}, {Orcinha}, {Palermo}, {Palmonari}, {Paniccia}, {Pashnin},
  {Pauluzzi}, {Pensotti}, {Perrina}, {Phan}, {Picot-Clemente}, {Plyaskin},
  {Pohl}, {Poireau}, {Popkow}, {Quadrani}, {Qi}, {Qin}, {Qu}, {Rancoita},
  {Rapin}, {Conde}, {Rosier-Lees}, {Rozhkov}, {Rozza}, {Sagdeev}, {Solano},
  {Schael}, {Schmidt}, {von Dratzig}, {Schwering}, {Seo}, {Shan}, {Shi},
  {Siedenburg}, {Song}, {Sun}, {Tacconi}, {Tang}, {Tang}, {Tian}, {Ting},
  {Ting}, {Tomassetti}, {Torsti}, {Urban}, {Vagelli}, {Valente}, {Valtonen},
  {Acosta}, {Vecchi}, {Velasco}, {Vialle}, {Viz{\'a}n}, {Wang}, {Wang}, {Wang},
  {Wang}, {Wang}, {Wang}, {Wei}, {Weng}, {Wu}, {Xiong}, {Xu}, {Yan}, {Yang},
  {Yi}, {Yu}, {Yu}, {Zannoni}, {Zeissler}, {Zhang}, {Zhang}, {Zhang}, {Zhang},
  {Zhao}, {Zheng}, {Zhuang}, {Zhukov}, {Zichichi}, {Zimmermann}, {Zuccon}, \&
  {AMS Collaboration}}]{2019PhRvL.122j1101A}
{Aguilar}, M., {Ali Cavasonza}, L., {Alpat}, B., {et~al.} 2019, \prl, 122,
  101101, \dodoi{10.1103/PhysRevLett.122.101101}

\bibitem[{{Alonso} {et~al.}(2015){Alonso}, {Bull}, {Ferreira}, \&
  {Santos}}]{2015MNRAS.447..400A}
{Alonso}, D., {Bull}, P., {Ferreira}, P.~G., \& {Santos}, M.~G. 2015, \mnras,
  447, 400, \dodoi{10.1093/mnras/stu2474}

\bibitem[{{Alonso} {et~al.}(2014){Alonso}, {Ferreira}, \&
  {Santos}}]{2014MNRAS.444.3183A}
{Alonso}, D., {Ferreira}, P.~G., \& {Santos}, M.~G. 2014, \mnras, 444, 3183,
  \dodoi{10.1093/mnras/stu1666}

\bibitem[{{Andreux} {et~al.}(2018){Andreux}, {Angles}, {Exarchakis},
  {Leonarduzzi}, {Rochette}, {Thiry}, {Zarka}, {Mallat}, {and{\'e}n},
  {Belilovsky}, {Bruna}, {Lostanlen}, {Chaudhary}, {Hirn}, {Oyallon}, {Zhang},
  {Cella}, \& {Eickenberg}}]{2018arXiv181211214A}
{Andreux}, M., {Angles}, T., {Exarchakis}, G., {et~al.} 2018, arXiv e-prints,
  arXiv:1812.11214, \dodoi{10.48550/arXiv.1812.11214}

\bibitem[{{Battye} {et~al.}(2013){Battye}, {Browne}, {Dickinson}, {Heron},
  {Maffei}, \& {Pourtsidou}}]{2013MNRAS.434.1239B}
{Battye}, R.~A., {Browne}, I.~W.~A., {Dickinson}, C., {et~al.} 2013, \mnras,
  434, 1239, \dodoi{10.1093/mnras/stt1082}

\bibitem[{{Battye} {et~al.}(2004){Battye}, {Davies}, \&
  {Weller}}]{2004MNRAS.355.1339B}
{Battye}, R.~A., {Davies}, R.~D., \& {Weller}, J. 2004, \mnras, 355, 1339,
  \dodoi{10.1111/j.1365-2966.2004.08416.x}

\bibitem[{{Bernardi} {et~al.}(2009){Bernardi}, {de Bruyn}, {Brentjens},
  {Ciardi}, {Harker}, {Jeli{\'c}}, {Koopmans}, {Labropoulos}, {Offringa},
  {Pandey}, {Schaye}, {Thomas}, {Yatawatta}, \&
  {Zaroubi}}]{2009A&A...500..965B}
{Bernardi}, G., {de Bruyn}, A.~G., {Brentjens}, M.~A., {et~al.} 2009, \aap,
  500, 965, \dodoi{10.1051/0004-6361/200911627}

\bibitem[{{Bernardi} {et~al.}(2010){Bernardi}, {de Bruyn}, {Harker},
  {Brentjens}, {Ciardi}, {Jeli{\'c}}, {Koopmans}, {Labropoulos}, {Offringa},
  {Pandey}, {Schaye}, {Thomas}, {Yatawatta}, \&
  {Zaroubi}}]{2010A&A...522A..67B}
{Bernardi}, G., {de Bruyn}, A.~G., {Harker}, G., {et~al.} 2010, \aap, 522, A67,
  \dodoi{10.1051/0004-6361/200913420}

\bibitem[{Betancourt(2017)}]{betancourt2017conceptual}
Betancourt, M. 2017, arXiv preprint arXiv:1701.02434

\bibitem[{Betancourt \& Girolami(2015)}]{betancourt2015hamiltonian}
Betancourt, M., \& Girolami, M. 2015, Current trends in Bayesian methodology
  with applications, 79, 2

\bibitem[{{Bharadwaj} {et~al.}(2001){Bharadwaj}, {Nath}, \&
  {Sethi}}]{2001JApA...22...21B}
{Bharadwaj}, S., {Nath}, B.~B., \& {Sethi}, S.~K. 2001, Journal of Astrophysics
  and Astronomy, 22, 21, \dodoi{10.1007/BF02933588}

\bibitem[{{Bhatnagar} \& {Nityananda}(2001)}]{2001A&A...375..344B}
{Bhatnagar}, S., \& {Nityananda}, R. 2001, \aap, 375, 344,
  \dodoi{10.1051/0004-6361:20010799}

\bibitem[{{Bigot-Sazy} {et~al.}(2015){Bigot-Sazy}, {Dickinson}, {Battye},
  {Browne}, {Ma}, {Maffei}, {Noviello}, {Remazeilles}, \&
  {Wilkinson}}]{2015MNRAS.454.3240B}
{Bigot-Sazy}, M.~A., {Dickinson}, C., {Battye}, R.~A., {et~al.} 2015, \mnras,
  454, 3240, \dodoi{10.1093/mnras/stv2153}

\bibitem[{Bingham {et~al.}(2019)Bingham, Chen, Jankowiak, Obermeyer, Pradhan,
  Karaletsos, Singh, Szerlip, Horsfall, \& Goodman}]{bingham2019pyro}
Bingham, E., Chen, J.~P., Jankowiak, M., {et~al.} 2019, J. Mach. Learn. Res.,
  20, 28:1.
\newblock \url{http://jmlr.org/papers/v20/18-403.html}

\bibitem[{Bradbury {et~al.}(2018)Bradbury, Frostig, Hawkins, Johnson, Leary,
  Maclaurin, Necula, Paszke, Vander{P}las, Wanderman-{M}ilne, \&
  Zhang}]{jax2018github}
Bradbury, J., Frostig, R., Hawkins, P., {et~al.} 2018, {JAX}: composable
  transformations of {P}ython+{N}um{P}y programs, 0.3.13.
\newblock \url{http://github.com/google/jax}

\bibitem[{{Carucci} {et~al.}(2020){Carucci}, {Irfan}, \&
  {Bobin}}]{2020MNRAS.499..304C}
{Carucci}, I.~P., {Irfan}, M.~O., \& {Bobin}, J. 2020, \mnras, 499, 304,
  \dodoi{10.1093/mnras/staa2854}

\bibitem[{{Chang} {et~al.}(2008){Chang}, {Pen}, {Peterson}, \&
  {McDonald}}]{2008PhRvL.100i1303C}
{Chang}, T.-C., {Pen}, U.-L., {Peterson}, J.~B., \& {McDonald}, P. 2008, \prl,
  100, 091303, \dodoi{10.1103/PhysRevLett.100.091303}

\bibitem[{{Chapman} {et~al.}(2012){Chapman}, {Abdalla}, {Harker}, {Jeli{\'c}},
  {Labropoulos}, {Zaroubi}, {Brentjens}, {de Bruyn}, \&
  {Koopmans}}]{2012MNRAS.423.2518C}
{Chapman}, E., {Abdalla}, F.~B., {Harker}, G., {et~al.} 2012, \mnras, 423,
  2518, \dodoi{10.1111/j.1365-2966.2012.21065.x}

\bibitem[{{Chapman} {et~al.}(2013){Chapman}, {Abdalla}, {Bobin}, {Starck},
  {Harker}, {Jeli{\'c}}, {Labropoulos}, {Zaroubi}, {Brentjens}, {de Bruyn}, \&
  {Koopmans}}]{2013MNRAS.429..165C}
{Chapman}, E., {Abdalla}, F.~B., {Bobin}, J., {et~al.} 2013, \mnras, 429, 165,
  \dodoi{10.1093/mnras/sts333}

\bibitem[{{Chen}(2012)}]{2012IJMPS..12..256C}
{Chen}, X. 2012, in International Journal of Modern Physics Conference Series,
  Vol.~12, International Journal of Modern Physics Conference Series, 256--263,
  \dodoi{10.1142/S2010194512006459}

\bibitem[{{Chen} {et~al.}(2023){Chen}, {Chapman}, {Wolz}, \&
  {Mazumder}}]{2023MNRAS.524.3724C}
{Chen}, Z., {Chapman}, E., {Wolz}, L., \& {Mazumder}, A. 2023, \mnras, 524,
  3724, \dodoi{10.1093/mnras/stad2102}

\bibitem[{{Cheng} \& {M{\'e}nard}(2021)}]{2021arXiv211201288C}
{Cheng}, S., \& {M{\'e}nard}, B. 2021, arXiv e-prints, arXiv:2112.01288,
  \dodoi{10.48550/arXiv.2112.01288}

\bibitem[{{Cheng} {et~al.}(2020){Cheng}, {Ting}, {M{\'e}nard}, \&
  {Bruna}}]{2020MNRAS.499.5902C}
{Cheng}, S., {Ting}, Y.-S., {M{\'e}nard}, B., \& {Bruna}, J. 2020, \mnras, 499,
  5902, \dodoi{10.1093/mnras/staa3165}

\bibitem[{{CHIME Collaboration} {et~al.}(2022){CHIME Collaboration}, {Amiri},
  {Bandura}, {Boskovic}, {Chen}, {Cliche}, {Deng}, {Denman}, {Dobbs},
  {Fandino}, {Foreman}, {Halpern}, {Hanna}, {Hill}, {Hinshaw}, {H{\"o}fer},
  {Kania}, {Klages}, {Landecker}, {MacEachern}, {Masui}, {Mena-Parra},
  {Milutinovic}, {Mirhosseini}, {Newburgh}, {Nitsche}, {Ordog}, {Pen},
  {Pinsonneault-Marotte}, {Polzin}, {Reda}, {Renard}, {Shaw}, {Siegel},
  {Singh}, {Smegal}, {Tretyakov}, {van Gassen}, {Vanderlinde}, {Wang}, {Wiebe},
  {Willis}, \& {Wulf}}]{2022ApJS..261...29C}
{CHIME Collaboration}, {Amiri}, M., {Bandura}, K., {et~al.} 2022, \apjs, 261,
  29, \dodoi{10.3847/1538-4365/ac6fd9}

\bibitem[{{Chluba} {et~al.}(2017){Chluba}, {Hill}, \&
  {Abitbol}}]{2017MNRAS.472.1195C}
{Chluba}, J., {Hill}, J.~C., \& {Abitbol}, M.~H. 2017, \mnras, 472, 1195,
  \dodoi{10.1093/mnras/stx1982}

\bibitem[{Cunnington {et~al.}(2021)Cunnington, Irfan, Carucci, Pourtsidou, \&
  Bobin}]{Cunnington2021}
Cunnington, S., Irfan, M.~O., Carucci, I.~P., Pourtsidou, A., \& Bobin, J.
  2021, \mnras, 504, 208, \dodoi{10.1093/mnras/stab856}

\bibitem[{{Cunnington} {et~al.}(2023){Cunnington}, {Li}, {Santos}, {Wang},
  {Carucci}, {Irfan}, {Pourtsidou}, {Spinelli}, {Wolz}, {Soares}, {Blake},
  {Bull}, {Engelbrecht}, {Fonseca}, {Grainge}, \& {Ma}}]{2023MNRAS.518.6262C}
{Cunnington}, S., {Li}, Y., {Santos}, M.~G., {et~al.} 2023, \mnras, 518, 6262,
  \dodoi{10.1093/mnras/stac3060}

\bibitem[{{Dai} \& {Seljak}(2023)}]{2023arXiv230604689D}
{Dai}, B., \& {Seljak}, U. 2023, arXiv e-prints, arXiv:2306.04689,
  \dodoi{10.48550/arXiv.2306.04689}

\bibitem[{Dewdney {et~al.}(2019)}]{dewdney2019ska1}
Dewdney, P., {et~al.} 2019, SKA1 design baseline description, Tech. rep.,
  SKA-TEL-SKO-0001075, internal SKA document

\bibitem[{{Di Matteo} {et~al.}(2002){Di Matteo}, {Perna}, {Abel}, \&
  {Rees}}]{2002ApJ...564..576D}
{Di Matteo}, T., {Perna}, R., {Abel}, T., \& {Rees}, M.~J. 2002, \apj, 564,
  576, \dodoi{10.1086/324293}

\bibitem[{{Diao} {et~al.}(2024){Diao}, {Li}, {Grumitt}, \&
  {Mao}}]{2024arXiv241001136D}
{Diao}, K., {Li}, Z., {Grumitt}, R. D.~P., \& {Mao}, Y. 2024, arXiv e-prints,
  arXiv:2410.01136, \dodoi{10.48550/arXiv.2410.01136}

\bibitem[{{Dickinson} {et~al.}(2003){Dickinson}, {Davies}, \&
  {Davis}}]{2003MNRAS.341..369D}
{Dickinson}, C., {Davies}, R.~D., \& {Davis}, R.~J. 2003, \mnras, 341, 369,
  \dodoi{10.1046/j.1365-8711.2003.06439.x}

\bibitem[{{Eickenberg} {et~al.}(2018){Eickenberg}, {Exarchakis}, {Hirn},
  {Mallat}, \& {Thiry}}]{2018JChPh.148x1732E}
{Eickenberg}, M., {Exarchakis}, G., {Hirn}, M., {Mallat}, S., \& {Thiry}, L.
  2018, \jcp, 148, 241732, \dodoi{10.1063/1.5023798}

\bibitem[{Flaxman {et~al.}(2015)Flaxman, Gelman, Neill, Smola, Vehtari, \&
  Wilson}]{flaxman2015fast}
Flaxman, S., Gelman, A., Neill, D., {et~al.} 2015, Manuscript in preparation

\bibitem[{Furlanetto(2016)}]{Furlanetto2016}
Furlanetto, S.~R. 2016, The 21-cm Line as a Probe of Reionization, ed.
  A.~Mesinger (Cham: Springer International Publishing), 247--280,
  \dodoi{10.1007/978-3-319-21957-8_9}

\bibitem[{Gabry {et~al.}(2019)Gabry, Simpson, Vehtari, Betancourt, \&
  Gelman}]{gabry2019visualization}
Gabry, J., Simpson, D., Vehtari, A., Betancourt, M., \& Gelman, A. 2019,
  Journal of the Royal Statistical Society Series A: Statistics in Society,
  182, 389

\bibitem[{{Gehlot} {et~al.}(2019){Gehlot}, {Mertens}, {Koopmans}, {Brentjens},
  {Zaroubi}, {Ciardi}, {Ghosh}, {Hatef}, {Iliev}, {Jeli{\'c}}, {}, {Kooistra},
  {Krause}, {Mellema}, {Mevius}, {Mitra}, {Offringa}, {Pandey}, {Sardarabadi},
  {Schaye}, {Silva}, {Vedantham}, \& {Yatawatta}}]{2019MNRAS.488.4271G}
{Gehlot}, B.~K., {Mertens}, F.~G., {Koopmans}, L.~V.~E., {et~al.} 2019, \mnras,
  488, 4271, \dodoi{10.1093/mnras/stz1937}

\bibitem[{Gelman(2006)}]{doi:10.1198/004017005000000661}
Gelman, A. 2006, Technometrics, 48, 432, \dodoi{10.1198/004017005000000661}

\bibitem[{Gelman {et~al.}(2013)Gelman, Carlin, Stern, Dunson, Vehtari, \&
  Rubin}]{gelman2013bayesian}
Gelman, A., Carlin, J.~B., Stern, H.~S., {et~al.} 2013, Bayesian data analysis,
  3rd edn., Chapman \& Hall/CRC Texts in Statistical Science Series (Boca
  Raton, Florida: CRC).
\newblock
  \url{https://www.worldcat.org/title/bayesian-data-analysis/oclc/966614951?referer=br&ht=edition}

\bibitem[{Gelman \& Hennig(2017)}]{gelman2017beyond}
Gelman, A., \& Hennig, C. 2017, Journal of the Royal Statistical Society Series
  A: Statistics in Society, 180, 967

\bibitem[{Gelman \& Hill(2006)}]{gelman2006data}
Gelman, A., \& Hill, J. 2006, Data analysis using regression and
  multilevel/hierarchical models (Cambridge university press)

\bibitem[{{Gelman} \& {Rubin}(1992)}]{1992StaSc...7..457G}
{Gelman}, A., \& {Rubin}, D.~B. 1992, Statistical Science, 7, 457,
  \dodoi{10.1214/ss/1177011136}

\bibitem[{{Gelman} {et~al.}(2017){Gelman}, {Simpson}, \&
  {Betancourt}}]{2017Entrp..19..555G}
{Gelman}, A., {Simpson}, D., \& {Betancourt}, M. 2017, Entropy, 19, 555,
  \dodoi{10.3390/e19100555}

\bibitem[{{Ghosh} {et~al.}(2020){Ghosh}, {Mertens}, {Bernardi}, {Santos},
  {Kern}, {Carilli}, {Grobler}, {Koopmans}, {Jacobs}, {Liu}, {Parsons},
  {Morales}, {Aguirre}, {Dillon}, {Hazelton}, {Smirnov}, {Gehlot}, {Matika},
  {Alexander}, {Ali}, {Beardsley}, {Benefo}, {Billings}, {Bowman}, {Bradley},
  {Cheng}, {Chichura}, {DeBoer}, {de Lera Acedo}, {Ewall-Wice}, {Fadana},
  {Fagnoni}, {Fortino}, {Fritz}, {Furlanetto}, {Gallardo}, {Glendenning},
  {Gorthi}, {Greig}, {Grobbelaar}, {Hickish}, {Josaitis}, {Julius}, {Igarashi},
  {Kariseb}, {Kohn}, {Kolopanis}, {Lekalake}, {Loots}, {MacMahon}, {Malan},
  {Malgas}, {Maree}, {Martinot}, {Mathison}, {Matsetela}, {Mesinger}, {Neben},
  {Nikolic}, {Nunhokee}, {Patra}, {Pieterse}, {Razavi-Ghods}, {Ringuette},
  {Robnett}, {Rosie}, {Sell}, {Smith}, {Syce}, {Tegmark}, {Thyagarajan},
  {Williams}, \& {Zheng}}]{2020MNRAS.495.2813G}
{Ghosh}, A., {Mertens}, F., {Bernardi}, G., {et~al.} 2020, \mnras, 495, 2813,
  \dodoi{10.1093/mnras/staa1331}

\bibitem[{{G{\'o}rski} {et~al.}(2005){G{\'o}rski}, {Hivon}, {Banday},
  {Wandelt}, {Hansen}, {Reinecke}, \& {Bartelmann}}]{2005ApJ...622..759G}
{G{\'o}rski}, K.~M., {Hivon}, E., {Banday}, A.~J., {et~al.} 2005, \apj, 622,
  759, \dodoi{10.1086/427976}

\bibitem[{{Greig} \& {Mesinger}(2017)}]{2017MNRAS.472.2651G}
{Greig}, B., \& {Mesinger}, A. 2017, \mnras, 472, 2651,
  \dodoi{10.1093/mnras/stx2118}

\bibitem[{{Greig} {et~al.}(2023){Greig}, {Ting}, \&
  {Kaurov}}]{2023MNRAS.519.5288G}
{Greig}, B., {Ting}, Y.-S., \& {Kaurov}, A.~A. 2023, \mnras, 519, 5288,
  \dodoi{10.1093/mnras/stac3822}

\bibitem[{{Grumitt} {et~al.}(2020){Grumitt}, {Jew}, \&
  {Dickinson}}]{2020MNRAS.496.4383G}
{Grumitt}, R.~D.~P., {Jew}, L. R.~P., \& {Dickinson}, C. 2020, \mnras, 496,
  4383, \dodoi{10.1093/mnras/staa1857}

\bibitem[{{Haslam} {et~al.}(1981){Haslam}, {Klein}, {Salter}, {Stoffel},
  {Wilson}, {Cleary}, {Cooke}, \& {Thomasson}}]{1981A&A...100..209H}
{Haslam}, C.~G.~T., {Klein}, U., {Salter}, C.~J., {et~al.} 1981, \aap, 100, 209

\bibitem[{{Haslam} {et~al.}(1982){Haslam}, {Salter}, {Stoffel}, \&
  {Wilson}}]{1982A&AS...47....1H}
{Haslam}, C.~G.~T., {Salter}, C.~J., {Stoffel}, H., \& {Wilson}, W.~E. 1982,
  \aaps, 47, 1

\bibitem[{Hoffman {et~al.}(2019)Hoffman, Sountsov, Dillon, Langmore, Tran, \&
  Vasudevan}]{hoffman2019neutra}
Hoffman, M., Sountsov, P., Dillon, J.~V., {et~al.} 2019, arXiv preprint
  arXiv:1903.03704

\bibitem[{Hoffman {et~al.}(2014)Hoffman, Gelman, {et~al.}}]{hoffman2014no}
Hoffman, M.~D., Gelman, A., {et~al.} 2014, J. Mach. Learn. Res., 15, 1593

\bibitem[{{Hopkins} {et~al.}(2014){Hopkins}, {Kere{\v{s}}}, {O{\~n}orbe},
  {Faucher-Gigu{\`e}re}, {Quataert}, {Murray}, \&
  {Bullock}}]{2014MNRAS.445..581H}
{Hopkins}, P.~F., {Kere{\v{s}}}, D., {O{\~n}orbe}, J., {et~al.} 2014, \mnras,
  445, 581, \dodoi{10.1093/mnras/stu1738}

\bibitem[{{Hothi} {et~al.}(2021){Hothi}, {Chapman}, {Pritchard}, {Mertens},
  {Koopmans}, {Ciardi}, {Gehlot}, {Ghara}, {Ghosh}, {Giri}, {Iliev},
  {Jeli{\'c}}, \& {Zaroubi}}]{2021MNRAS.500.2264H}
{Hothi}, I., {Chapman}, E., {Pritchard}, J.~R., {et~al.} 2021, \mnras, 500,
  2264, \dodoi{10.1093/mnras/staa3446}

\bibitem[{{Irfan} {et~al.}(2022){Irfan}, {Bull}, {Santos}, {Wang}, {Grainge},
  {Li}, {Carucci}, {Spinelli}, \& {Cunnington}}]{2022MNRAS.509.4923I}
{Irfan}, M.~O., {Bull}, P., {Santos}, M.~G., {et~al.} 2022, \mnras, 509, 4923,
  \dodoi{10.1093/mnras/stab3346}

\bibitem[{{Jeli{\'c}} {et~al.}(2008){Jeli{\'c}}, {Zaroubi}, {Labropoulos},
  {Thomas}, {Bernardi}, {Brentjens}, {de Bruyn}, {Ciardi}, {Harker},
  {Koopmans}, {Pandey}, {Schaye}, \& {Yatawatta}}]{2008MNRAS.389.1319J}
{Jeli{\'c}}, V., {Zaroubi}, S., {Labropoulos}, P., {et~al.} 2008, \mnras, 389,
  1319, \dodoi{10.1111/j.1365-2966.2008.13634.x}

\bibitem[{{Jew} \& {Grumitt}(2020)}]{2020MNRAS.495..578J}
{Jew}, L., \& {Grumitt}, R.~D.~P. 2020, \mnras, 495, 578,
  \dodoi{10.1093/mnras/staa1233}

\bibitem[{{Kern} \& {Liu}(2021)}]{2021MNRAS.501.1463K}
{Kern}, N.~S., \& {Liu}, A. 2021, \mnras, 501, 1463,
  \dodoi{10.1093/mnras/staa3736}

\bibitem[{Klypin {et~al.}(2016)Klypin, Yepes, Gottlöber, Prada, \&
  Heß}]{Klypin2016}
Klypin, A., Yepes, G., Gottlöber, S., Prada, F., \& Heß, S. 2016, MNRAS, 457,
  4340

\bibitem[{Kumar {et~al.}(2019)Kumar, Carroll, Hartikainen, \&
  Martin}]{Kumar2019}
Kumar, R., Carroll, C., Hartikainen, A., \& Martin, O. 2019, Journal of Open
  Source Software, 4, 1143, \dodoi{10.21105/joss.01143}

\bibitem[{Lalchand \& Rasmussen(2020)}]{lalchand2020approximate}
Lalchand, V., \& Rasmussen, C.~E. 2020, in Symposium on Advances in Approximate
  Bayesian Inference, PMLR, 1--12

\bibitem[{{Liu} \& {Shaw}(2020)}]{2020PASP..132f2001L}
{Liu}, A., \& {Shaw}, J.~R. 2020, \pasp, 132, 062001,
  \dodoi{10.1088/1538-3873/ab5bfd}

\bibitem[{{Liu} \& {Tegmark}(2011)}]{2011PhRvD..83j3006L}
{Liu}, A., \& {Tegmark}, M. 2011, \prd, 83, 103006,
  \dodoi{10.1103/PhysRevD.83.103006}

\bibitem[{{Liu} \& {Tegmark}(2012)}]{2012MNRAS.419.3491L}
---. 2012, \mnras, 419, 3491, \dodoi{10.1111/j.1365-2966.2011.19989.x}

\bibitem[{Livingstone {et~al.}(2016)Livingstone, Betancourt, Byrne, \&
  Girolami}]{livingstone_gergo2016}
Livingstone, S., Betancourt, M., Byrne, S., \& Girolami, M. 2016, Bernoulli,
  25, \dodoi{10.3150/18-BEJ1083}

\bibitem[{MacKay(2003)}]{mackay2003information}
MacKay, D. 2003, Information Theory, Inference and Learning Algorithms
  (Cambridge University Press).
\newblock \url{https://books.google.co.uk/books?id=AKuMj4PN_EMC}

\bibitem[{{Masui} {et~al.}(2013){Masui}, {Switzer}, {Banavar}, {Bandura},
  {Blake}, {Calin}, {Chang}, {Chen}, {Li}, {Liao}, {Natarajan}, {Pen},
  {Peterson}, {Shaw}, \& {Voytek}}]{2013ApJ...763L..20M}
{Masui}, K.~W., {Switzer}, E.~R., {Banavar}, N., {et~al.} 2013, \apjl, 763,
  L20, \dodoi{10.1088/2041-8205/763/1/L20}

\bibitem[{{MeerKLASS Collaboration} {et~al.}(2024){MeerKLASS Collaboration},
  {Barberi-Squarotti}, {Bernal}, {Bull}, {Camera}, {Carucci}, {Chen},
  {Cunnington}, {Engelbrecht}, {Fonseca}, {Grainge}, {Irfan}, {Li}, {Mazumder},
  {Paul}, {Pourtsidou}, {Santos}, {Spinelli}, {Wang}, {Witzemann}, \&
  {Wolz}}]{2024arXiv240721626M}
{MeerKLASS Collaboration}, {Barberi-Squarotti}, M., {Bernal}, J.~L., {et~al.}
  2024, arXiv e-prints, arXiv:2407.21626, \dodoi{10.48550/arXiv.2407.21626}

\bibitem[{{Mertens} {et~al.}(2024){Mertens}, {Bobin}, \&
  {Carucci}}]{2024MNRAS.527.3517M}
{Mertens}, F.~G., {Bobin}, J., \& {Carucci}, I.~P. 2024, \mnras, 527, 3517,
  \dodoi{10.1093/mnras/stad3430}

\bibitem[{{Mertens} {et~al.}(2018){Mertens}, {Ghosh}, \&
  {Koopmans}}]{2018MNRAS.478.3640M}
{Mertens}, F.~G., {Ghosh}, A., \& {Koopmans}, L.~V.~E. 2018, \mnras, 478, 3640,
  \dodoi{10.1093/mnras/sty1207}

\bibitem[{{Mertens} {et~al.}(2020){Mertens}, {Mevius}, {Koopmans}, {Offringa},
  {Mellema}, {Zaroubi}, {Brentjens}, {Gan}, {Gehlot}, {Pandey}, {Sardarabadi},
  {Vedantham}, {Yatawatta}, {Asad}, {Ciardi}, {Chapman}, {Gazagnes}, {Ghara},
  {Ghosh}, {Giri}, {Iliev}, {Jeli{\'c}}, {Kooistra}, {Mondal}, {Schaye}, \&
  {Silva}}]{2020MNRAS.493.1662M}
{Mertens}, F.~G., {Mevius}, M., {Koopmans}, L.~V.~E., {et~al.} 2020, \mnras,
  493, 1662, \dodoi{10.1093/mnras/staa327}

\bibitem[{{Nunhokee} {et~al.}(2017){Nunhokee}, {Bernardi}, {Kohn}, {Aguirre},
  {Thyagarajan}, {Dillon}, {Foster}, {Grobler}, {Martinot}, \&
  {Parsons}}]{2017ApJ...848...47N}
{Nunhokee}, C.~D., {Bernardi}, G., {Kohn}, S.~A., {et~al.} 2017, \apj, 848, 47,
  \dodoi{10.3847/1538-4357/aa8b73}

\bibitem[{{Offringa} {et~al.}(2019){Offringa}, {Mertens}, \&
  {Koopmans}}]{2019MNRAS.484.2866O}
{Offringa}, A.~R., {Mertens}, F., \& {Koopmans}, L.~V.~E. 2019, \mnras, 484,
  2866, \dodoi{10.1093/mnras/stz175}

\bibitem[{{Oh} \& {Mack}(2003)}]{2003MNRAS.346..871O}
{Oh}, S.~P., \& {Mack}, K.~J. 2003, \mnras, 346, 871,
  \dodoi{10.1111/j.1365-2966.2003.07133.x}

\bibitem[{{Olivari} {et~al.}(2018){Olivari}, {Dickinson}, {Battye}, {Ma},
  {Costa}, {Remazeilles}, \& {Harper}}]{2018MNRAS.473.4242O}
{Olivari}, L.~C., {Dickinson}, C., {Battye}, R.~A., {et~al.} 2018, \mnras, 473,
  4242, \dodoi{10.1093/mnras/stx2621}

\bibitem[{{Paul} {et~al.}(2023){Paul}, {Santos}, {Chen}, \&
  {Wolz}}]{2023arXiv230111943P}
{Paul}, S., {Santos}, M.~G., {Chen}, Z., \& {Wolz}, L. 2023, arXiv e-prints,
  arXiv:2301.11943, \dodoi{10.48550/arXiv.2301.11943}

\bibitem[{Phan {et~al.}(2019)Phan, Pradhan, \& Jankowiak}]{phan2019composable}
Phan, D., Pradhan, N., \& Jankowiak, M. 2019, arXiv preprint arXiv:1912.11554

\bibitem[{{Planck Collaboration} {et~al.}(2016{\natexlab{a}}){Planck
  Collaboration}, {Adam}, {Ade}, {Aghanim}, {Alves}, {Arnaud}, {Ashdown},
  {Aumont}, {Baccigalupi}, {Banday}, {Barreiro}, {Bartlett}, {Bartolo},
  {Battaner}, {Benabed}, {Beno{\^\i}t}, {Benoit-L{\'e}vy}, {Bernard},
  {Bersanelli}, {Bielewicz}, {Bock}, {Bonaldi}, {Bonavera}, {Bond}, {Borrill},
  {Bouchet}, {Boulanger}, {Bucher}, {Burigana}, {Butler}, {Calabrese},
  {Cardoso}, {Catalano}, {Challinor}, {Chamballu}, {Chary}, {Chiang},
  {Christensen}, {Clements}, {Colombi}, {Colombo}, {Combet}, {Couchot},
  {Coulais}, {Crill}, {Curto}, {Cuttaia}, {Danese}, {Davies}, {Davis}, {de
  Bernardis}, {de Rosa}, {de Zotti}, {Delabrouille}, {D{\'e}sert}, {Dickinson},
  {Diego}, {Dole}, {Donzelli}, {Dor{\'e}}, {Douspis}, {Ducout}, {Dupac},
  {Efstathiou}, {Elsner}, {En{\ss}lin}, {Eriksen}, {Falgarone}, {Fergusson},
  {Finelli}, {Forni}, {Frailis}, {Fraisse}, {Franceschi}, {Frejsel},
  {Galeotta}, {Galli}, {Ganga}, {Ghosh}, {Giard}, {Giraud-H{\'e}raud},
  {Gjerl{\o}w}, {Gonz{\'a}lez-Nuevo}, {G{\'o}rski}, {Gratton}, {Gregorio},
  {Gruppuso}, {Gudmundsson}, {Hansen}, {Hanson}, {Harrison}, {Helou},
  {Henrot-Versill{\'e}}, {Hern{\'a}ndez-Monteagudo}, {Herranz}, {Hildebrandt},
  {Hivon}, {Hobson}, {Holmes}, {Hornstrup}, {Hovest}, {Huffenberger}, {Hurier},
  {Jaffe}, {Jaffe}, {Jones}, {Juvela}, {Keih{\"a}nen}, {Keskitalo}, {Kisner},
  {Kneissl}, {Knoche}, {Kunz}, {Kurki-Suonio}, {Lagache},
  {L{\"a}hteenm{\"a}ki}, {Lamarre}, {Lasenby}, {Lattanzi}, {Lawrence}, {Le
  Jeune}, {Leahy}, {Leonardi}, {Lesgourgues}, {Levrier}, {Liguori}, {Lilje},
  {Linden-V{\o}rnle}, {L{\'o}pez-Caniego}, {Lubin}, {Mac{\'\i}as-P{\'e}rez},
  {Maggio}, {Maino}, {Mandolesi}, {Mangilli}, {Maris}, {Marshall}, {Martin},
  {Mart{\'\i}nez-Gonz{\'a}lez}, {Masi}, {Matarrese}, {McGehee}, {Meinhold},
  {Melchiorri}, {Mendes}, {Mennella}, {Migliaccio}, {Mitra},
  {Miville-Desch{\^e}nes}, {Moneti}, {Montier}, {Morgante}, {Mortlock}, {Moss},
  {Munshi}, {Murphy}, {Naselsky}, {Nati}, {Natoli}, {Netterfield},
  {N{\o}rgaard-Nielsen}, {Noviello}, {Novikov}, {Novikov}, {Orlando},
  {Oxborrow}, {Paci}, {Pagano}, {Pajot}, {Paladini}, {Paoletti}, {Partridge},
  {Pasian}, {Patanchon}, {Pearson}, {Perdereau}, {Perotto}, {Perrotta},
  {Pettorino}, {Piacentini}, {Piat}, {Pierpaoli}, {Pietrobon}, {Plaszczynski},
  {Pointecouteau}, {Polenta}, {Pratt}, {Pr{\'e}zeau}, {Prunet}, {Puget},
  {Rachen}, {Reach}, {Rebolo}, {Reinecke}, {Remazeilles}, {Renault}, {Renzi},
  {Ristorcelli}, {Rocha}, {Rosset}, {Rossetti}, {Roudier},
  {Rubi{\~n}o-Mart{\'\i}n}, {Rusholme}, {Sandri}, {Santos}, {Savelainen},
  {Savini}, {Scott}, {Seiffert}, {Shellard}, {Spencer}, {Stolyarov}, {Stompor},
  {Strong}, {Sudiwala}, {Sunyaev}, {Sutton}, {Suur-Uski}, {Sygnet}, {Tauber},
  {Terenzi}, {Toffolatti}, {Tomasi}, {Tristram}, {Tucci}, {Tuovinen}, {Umana},
  {Valenziano}, {Valiviita}, {Van Tent}, {Vielva}, {Villa}, {Wade}, {Wandelt},
  {Wehus}, {Wilkinson}, {Yvon}, {Zacchei}, \& {Zonca}}]{2016A&A...594A..10P}
{Planck Collaboration}, {Adam}, R., {Ade}, P.~A.~R., {et~al.}
  2016{\natexlab{a}}, \aap, 594, A10, \dodoi{10.1051/0004-6361/201525967}

\bibitem[{{Planck Collaboration} {et~al.}(2016{\natexlab{b}}){Planck
  Collaboration}, {Ade}, {Aghanim}, {Arnaud}, {Ashdown}, {Aumont},
  {Baccigalupi}, {Banday}, {Barreiro}, {Bartlett}, {Bartolo}, {Battaner},
  {Battye}, {Benabed}, {Beno{\^\i}t}, {Benoit-L{\'e}vy}, {Bernard},
  {Bersanelli}, {Bielewicz}, {Bock}, {Bonaldi}, {Bonavera}, {Bond}, {Borrill},
  {Bouchet}, {Boulanger}, {Bucher}, {Burigana}, {Butler}, {Calabrese},
  {Cardoso}, {Catalano}, {Challinor}, {Chamballu}, {Chary}, {Chiang}, {Chluba},
  {Christensen}, {Church}, {Clements}, {Colombi}, {Colombo}, {Combet},
  {Coulais}, {Crill}, {Curto}, {Cuttaia}, {Danese}, {Davies}, {Davis}, {de
  Bernardis}, {de Rosa}, {de Zotti}, {Delabrouille}, {D{\'e}sert}, {Di
  Valentino}, {Dickinson}, {Diego}, {Dolag}, {Dole}, {Donzelli}, {Dor{\'e}},
  {Douspis}, {Ducout}, {Dunkley}, {Dupac}, {Efstathiou}, {Elsner},
  {En{\ss}lin}, {Eriksen}, {Farhang}, {Fergusson}, {Finelli}, {Forni},
  {Frailis}, {Fraisse}, {Franceschi}, {Frejsel}, {Galeotta}, {Galli}, {Ganga},
  {Gauthier}, {Gerbino}, {Ghosh}, {Giard}, {Giraud-H{\'e}raud}, {Giusarma},
  {Gjerl{\o}w}, {Gonz{\'a}lez-Nuevo}, {G{\'o}rski}, {Gratton}, {Gregorio},
  {Gruppuso}, {Gudmundsson}, {Hamann}, {Hansen}, {Hanson}, {Harrison}, {Helou},
  {Henrot-Versill{\'e}}, {Hern{\'a}ndez-Monteagudo}, {Herranz}, {Hildebrandt},
  {Hivon}, {Hobson}, {Holmes}, {Hornstrup}, {Hovest}, {Huang}, {Huffenberger},
  {Hurier}, {Jaffe}, {Jaffe}, {Jones}, {Juvela}, {Keih{\"a}nen}, {Keskitalo},
  {Kisner}, {Kneissl}, {Knoche}, {Knox}, {Kunz}, {Kurki-Suonio}, {Lagache},
  {L{\"a}hteenm{\"a}ki}, {Lamarre}, {Lasenby}, {Lattanzi}, {Lawrence}, {Leahy},
  {Leonardi}, {Lesgourgues}, {Levrier}, {Lewis}, {Liguori}, {Lilje},
  {Linden-V{\o}rnle}, {L{\'o}pez-Caniego}, {Lubin}, {Mac{\'\i}as-P{\'e}rez},
  {Maggio}, {Maino}, {Mandolesi}, {Mangilli}, {Marchini}, {Maris}, {Martin},
  {Martinelli}, {Mart{\'\i}nez-Gonz{\'a}lez}, {Masi}, {Matarrese}, {McGehee},
  {Meinhold}, {Melchiorri}, {Melin}, {Mendes}, {Mennella}, {Migliaccio},
  {Millea}, {Mitra}, {Miville-Desch{\^e}nes}, {Moneti}, {Montier}, {Morgante},
  {Mortlock}, {Moss}, {Munshi}, {Murphy}, {Naselsky}, {Nati}, {Natoli},
  {Netterfield}, {N{\o}rgaard-Nielsen}, {Noviello}, {Novikov}, {Novikov},
  {Oxborrow}, {Paci}, {Pagano}, {Pajot}, {Paladini}, {Paoletti}, {Partridge},
  {Pasian}, {Patanchon}, {Pearson}, {Perdereau}, {Perotto}, {Perrotta},
  {Pettorino}, {Piacentini}, {Piat}, {Pierpaoli}, {Pietrobon}, {Plaszczynski},
  {Pointecouteau}, {Polenta}, {Popa}, {Pratt}, {Pr{\'e}zeau}, {Prunet},
  {Puget}, {Rachen}, {Reach}, {Rebolo}, {Reinecke}, {Remazeilles}, {Renault},
  {Renzi}, {Ristorcelli}, {Rocha}, {Rosset}, {Rossetti}, {Roudier},
  {Rouill{\'e} d'Orfeuil}, {Rowan-Robinson}, {Rubi{\~n}o-Mart{\'\i}n},
  {Rusholme}, {Said}, {Salvatelli}, {Salvati}, {Sandri}, {Santos},
  {Savelainen}, {Savini}, {Scott}, {Seiffert}, {Serra}, {Shellard}, {Spencer},
  {Spinelli}, {Stolyarov}, {Stompor}, {Sudiwala}, {Sunyaev}, {Sutton},
  {Suur-Uski}, {Sygnet}, {Tauber}, {Terenzi}, {Toffolatti}, {Tomasi},
  {Tristram}, {Trombetti}, {Tucci}, {Tuovinen}, {T{\"u}rler}, {Umana},
  {Valenziano}, {Valiviita}, {Van Tent}, {Vielva}, {Villa}, {Wade}, {Wandelt},
  {Wehus}, {White}, {White}, {Wilkinson}, {Yvon}, {Zacchei}, \&
  {Zonca}}]{2016A&A...594A..13P}
{Planck Collaboration}, {Ade}, P.~A.~R., {Aghanim}, N., {et~al.}
  2016{\natexlab{b}}, \aap, 594, A13, \dodoi{10.1051/0004-6361/201525830}

\bibitem[{{Pritchard} \& {Loeb}(2012)}]{2012RPPh...75h6901P}
{Pritchard}, J.~R., \& {Loeb}, A. 2012, Reports on Progress in Physics, 75,
  086901, \dodoi{10.1088/0034-4885/75/8/086901}

\bibitem[{Rasmussen \& Williams(2006)}]{RasmussenW06}
Rasmussen, C.~E., \& Williams, C. K.~I. 2006, Gaussian processes for machine
  learning., Adaptive computation and machine learning (MIT Press), I--XVIII,
  1--248

\bibitem[{{Remazeilles} {et~al.}(2015){Remazeilles}, {Dickinson}, {Banday},
  {Bigot-Sazy}, \& {Ghosh}}]{2015MNRAS.451.4311R}
{Remazeilles}, M., {Dickinson}, C., {Banday}, A.~J., {Bigot-Sazy}, M.~A., \&
  {Ghosh}, T. 2015, \mnras, 451, 4311, \dodoi{10.1093/mnras/stv1274}

\bibitem[{{Remazeilles} {et~al.}(2018){Remazeilles}, {Banday}, {Baccigalupi},
  {Basak}, {Bonaldi}, {De Zotti}, {Delabrouille}, {Dickinson}, {Eriksen},
  {Errard}, {Fernandez-Cobos}, {Fuskeland}, {Herv{\'\i}as-Caimapo},
  {L{\'o}pez-Caniego}, {Martinez-Gonz{\'a}lez}, {Roman}, {Vielva}, {Wehus},
  {Achucarro}, {Ade}, {Allison}, {Ashdown}, {Ballardini}, {Banerji},
  {Bartlett}, {Bartolo}, {Baumann}, {Bersanelli}, {Bonato}, {Borrill},
  {Bouchet}, {Boulanger}, {Brinckmann}, {Bucher}, {Burigana}, {Buzzelli},
  {Cai}, {Calvo}, {Carvalho}, {Castellano}, {Challinor}, {Chluba}, {Clesse},
  {Colantoni}, {Coppolecchia}, {Crook}, {D'Alessandro}, {de Bernardis}, {de
  Gasperis}, {Diego}, {Di Valentino}, {Feeney}, {Ferraro}, {Finelli},
  {Forastieri}, {Galli}, {Genova-Santos}, {Gerbino}, {Gonz{\'a}lez-Nuevo},
  {Grandis}, {Greenslade}, {Hagstotz}, {Hanany}, {Handley},
  {Hernandez-Monteagudo}, {Hills}, {Hivon}, {Kiiveri}, {Kisner}, {Kitching},
  {Kunz}, {Kurki-Suonio}, {Lamagna}, {Lasenby}, {Lattanzi}, {Lesgourgues},
  {Lewis}, {Liguori}, {Lindholm}, {Luzzi}, {Maffei}, {Martins}, {Masi},
  {Matarrese}, {McCarthy}, {Melin}, {Melchiorri}, {Molinari}, {Monfardini},
  {Natoli}, {Negrello}, {Notari}, {Paiella}, {Paoletti}, {Patanchon}, {Piat},
  {Pisano}, {Polastri}, {Polenta}, {Pollo}, {Poulin}, {Quartin},
  {Rubino-Martin}, {Salvati}, {Tartari}, {Tomasi}, {Tramonte}, {Trappe},
  {Trombetti}, {Tucker}, {Valiviita}, {Van de Weijgaert}, {van Tent}, {Vennin},
  {Vittorio}, {Young}, \& {Zannoni}}]{2018JCAP...04..023R}
{Remazeilles}, M., {Banday}, A.~J., {Baccigalupi}, C., {et~al.} 2018, \jcap,
  2018, 023, \dodoi{10.1088/1475-7516/2018/04/023}

\bibitem[{Riutort-Mayol {et~al.}(2023)Riutort-Mayol, B{\"u}rkner, Andersen,
  Solin, \& Vehtari}]{riutort2023practical}
Riutort-Mayol, G., B{\"u}rkner, P.-C., Andersen, M.~R., Solin, A., \& Vehtari,
  A. 2023, Statistics and Computing, 33, 17

\bibitem[{{Robnik} {et~al.}(2022){Robnik}, {De Luca}, {Silverstein}, \&
  {Seljak}}]{2022arXiv221208549R}
{Robnik}, J., {De Luca}, G.~B., {Silverstein}, E., \& {Seljak}, U. 2022, arXiv
  e-prints, arXiv:2212.08549, \dodoi{10.48550/arXiv.2212.08549}

\bibitem[{{Ruiz-Zapatero} {et~al.}(2022){Ruiz-Zapatero},
  {Garc{\'\i}a-Garc{\'\i}a}, {Alonso}, {Ferreira}, \&
  {Grumitt}}]{2022MNRAS.512.1967R}
{Ruiz-Zapatero}, J., {Garc{\'\i}a-Garc{\'\i}a}, C., {Alonso}, D., {Ferreira},
  P.~G., \& {Grumitt}, R. D.~P. 2022, \mnras, 512, 1967,
  \dodoi{10.1093/mnras/stac431}

\bibitem[{{Santos} {et~al.}(2016){Santos}, {Bull}, {Camera}, {Chen}, {Fonseca},
  {Heywood}, {Hilton}, {Jarvis}, {Jozsa}, {Knowles}, {Leeuw}, {Maartens},
  {Malefahlo}, {McAlpine}, {Moodley}, {Patel}, {Pourtsidou}, {Prescott},
  {Spekkens}, {Taylor}, {Witzemann}, \& {Whittam}}]{2016mks..confE..32S}
{Santos}, M., {Bull}, P., {Camera}, S., {et~al.} 2016, in MeerKAT Science: On
  the Pathway to the SKA, 32, \dodoi{10.22323/1.277.0032}

\bibitem[{{Shafieloo} {et~al.}(2012){Shafieloo}, {Kim}, \&
  {Linder}}]{2012PhRvD..85l3530S}
{Shafieloo}, A., {Kim}, A.~G., \& {Linder}, E.~V. 2012, \prd, 85, 123530,
  \dodoi{10.1103/PhysRevD.85.123530}

\bibitem[{Shaw {et~al.}(2015)Shaw, Sigurdson, Sitwell, Stebbins, \&
  Pen}]{PhysRevD.91.083514}
Shaw, J.~R., Sigurdson, K., Sitwell, M., Stebbins, A., \& Pen, U.-L. 2015,
  Phys. Rev. D, 91, 083514, \dodoi{10.1103/PhysRevD.91.083514}

\bibitem[{{Shaw} {et~al.}(2015){Shaw}, {Sigurdson}, {Sitwell}, {Stebbins}, \&
  {Pen}}]{2015PhRvD..91h3514S}
{Shaw}, J.~R., {Sigurdson}, K., {Sitwell}, M., {Stebbins}, A., \& {Pen}, U.-L.
  2015, \prd, 91, 083514, \dodoi{10.1103/PhysRevD.91.083514}

\bibitem[{{Slosar} {et~al.}(2019){Slosar}, {Ahmed}, {Alonso}, {Amin}, {Arena},
  {Bandura}, {Battaglia}, {Blazek}, {Bull}, {Castorina}, {Chang}, {Connor},
  {Dav{\'e}}, {Dvorkin}, {van Engelen}, {Ferraro}, {Flauger}, {Foreman},
  {Frisch}, {Green}, {Holder}, {Jacobs}, {Johnson}, {Dillon}, {Karagiannis},
  {Kaurov}, {Knox}, {Liu}, {Loverde}, {Ma}, {Masui}, {McClintock}, {Moodley},
  {Munchmeyer}, {Newburgh}, {Ng}, {Nomerotski}, {O'Connor}, {Obuljen},
  {Padmanabhan}, {Parkinson}, {Prochaska}, {Rajendran}, {Rapetti},
  {Saliwanchik}, {Schaan}, {Sehgal}, {Shaw}, {Sheehy}, {Sheldon}, {Shirley},
  {Silverstein}, {Slatyer}, {Slosar}, {Stankus}, {Stebbins}, {Timbie},
  {Tucker}, {Tyndall}, {Villaescusa Navarro}, {Wallisch}, \&
  {White}}]{2019BAAS...51g..53S}
{Slosar}, A., {Ahmed}, Z., {Alonso}, D., {et~al.} 2019, in Bulletin of the
  American Astronomical Society, Vol.~51, 53, \dodoi{10.48550/arXiv.1907.12559}

\bibitem[{Soares {et~al.}(2022)Soares, Watkinson, Cunnington, \&
  Pourtsidou}]{Soares_2022}
Soares, P.~S., Watkinson, C.~A., Cunnington, S., \& Pourtsidou, A. 2022,
  \mnras, 510, 5872, \dodoi{10.1093/mnras/stab2594}

\bibitem[{Solin \& S{\"a}rkk{\"a}(2020)}]{solin2020hilbert}
Solin, A., \& S{\"a}rkk{\"a}, S. 2020, Statistics and Computing, 30, 419

\bibitem[{{Stan Development Team}(2012)}]{stan_development_team_stan_2012}
{Stan Development Team}. 2012, Stan {Modeling} {Language} {User}'s {Guide} and
  {Reference} {Manual}, {Version} 1.0.
\newblock \url{http://mc-stan.org/}

\bibitem[{{Sui} {et~al.}(2023){Sui}, {Zhao}, {Jing}, \&
  {Mao}}]{2023arXiv230704994S}
{Sui}, C., {Zhao}, X., {Jing}, T., \& {Mao}, Y. 2023, arXiv e-prints,
  arXiv:2307.04994, \dodoi{10.48550/arXiv.2307.04994}

\bibitem[{{Switzer} {et~al.}(2015){Switzer}, {Chang}, {Masui}, {Pen}, \&
  {Voytek}}]{2015ApJ...815...51S}
{Switzer}, E.~R., {Chang}, T.~C., {Masui}, K.~W., {Pen}, U.~L., \& {Voytek},
  T.~C. 2015, \apj, 815, 51, \dodoi{10.1088/0004-637X/815/1/51}

\bibitem[{{Vanderlinde} {et~al.}(2019){Vanderlinde}, {Liu}, {Gaensler}, {Bond},
  {Hinshaw}, {Ng}, {Chiang}, {Stairs}, {Brown}, {Sievers}, {Mena}, {Smith},
  {Bandura}, {Masui}, {Spekkens}, {Belostotski}, {Dobbs}, {Turok}, {Boyle},
  {Rupen}, {Landecker}, {Pen}, \& {Kaspi}}]{2019clrp.2020...28V}
{Vanderlinde}, K., {Liu}, A., {Gaensler}, B., {et~al.} 2019, in Canadian Long
  Range Plan for Astronomy and Astrophysics White Papers, Vol. 2020, 28,
  \dodoi{10.5281/zenodo.3765414}

\bibitem[{Vehtari {et~al.}(2017)Vehtari, Gelman, \&
  Gabry}]{vehtari2017practical}
Vehtari, A., Gelman, A., \& Gabry, J. 2017, Statistics and computing, 27, 1413

\bibitem[{{Vehtari} {et~al.}(2015){Vehtari}, {Simpson}, {Gelman}, {Yao}, \&
  {Gabry}}]{2015arXiv150702646V}
{Vehtari}, A., {Simpson}, D., {Gelman}, A., {Yao}, Y., \& {Gabry}, J. 2015,
  arXiv e-prints, arXiv:1507.02646, \dodoi{10.48550/arXiv.1507.02646}

\bibitem[{{Villaescusa-Navarro} {et~al.}(2018){Villaescusa-Navarro}, {Genel},
  {Castorina}, {Obuljen}, {Spergel}, {Hernquist}, {Nelson}, {Carucci},
  {Pillepich}, {Marinacci}, {Diemer}, {Vogelsberger}, {Weinberger}, \&
  {Pakmor}}]{2018ApJ...866..135V}
{Villaescusa-Navarro}, F., {Genel}, S., {Castorina}, E., {et~al.} 2018, \apj,
  866, 135, \dodoi{10.3847/1538-4357/aadba0}

\bibitem[{{Waelkens} {et~al.}(2009){Waelkens}, {Jaffe}, {Reinecke}, {Kitaura},
  \& {En{\ss}lin}}]{2009A&A...495..697W}
{Waelkens}, A., {Jaffe}, T., {Reinecke}, M., {Kitaura}, F.~S., \& {En{\ss}lin},
  T.~A. 2009, \aap, 495, 697, \dodoi{10.1051/0004-6361:200810564}

\bibitem[{{Wang} {et~al.}(2020){Wang}, {Jaffe}, {En{\ss}lin}, {Ullio}, {Ghosh},
  \& {Santos}}]{2020ApJS..247...18W}
{Wang}, J., {Jaffe}, T.~R., {En{\ss}lin}, T.~A., {et~al.} 2020, \apjs, 247, 18,
  \dodoi{10.3847/1538-4365/ab72a2}

\bibitem[{{Wang} {et~al.}(2006){Wang}, {Tegmark}, {Santos}, \&
  {Knox}}]{2006ApJ...650..529W}
{Wang}, X., {Tegmark}, M., {Santos}, M.~G., \& {Knox}, L. 2006, \apj, 650, 529,
  \dodoi{10.1086/506597}

\bibitem[{{Weltman} {et~al.}(2020){Weltman}, {Bull}, {Camera}, {Kelley},
  {Padmanabhan}, {Pritchard}, {Raccanelli}, {Riemer-S{\o}rensen}, {Shao},
  {Andrianomena}, {Athanassoula}, {Bacon}, {Barkana}, {Bertone}, {B{\oe}hm},
  {Bonvin}, {Bosma}, {Br{\"u}ggen}, {Burigana}, {Calore}, {Cembranos},
  {Clarkson}, {Connors}, {Cruz-Dombriz}, {Dunsby}, {Fonseca}, {Fornengo},
  {Gaggero}, {Harrison}, {Larena}, {Ma}, {Maartens}, {M{\'e}ndez-Isla},
  {Mohanty}, {Murray}, {Parkinson}, {Pourtsidou}, {Quinn}, {Regis}, {Saha},
  {Sahl{\'e}n}, {Sakellariadou}, {Silk}, {Trombetti}, {Vazza}, {Venumadhav},
  {Vidotto}, {Villaescusa-Navarro}, {Wang}, {Weniger}, {Wolz}, {Zhang}, \&
  {Gaensler}}]{2020PASA...37....2W}
{Weltman}, A., {Bull}, P., {Camera}, S., {et~al.} 2020, \pasa, 37, e002,
  \dodoi{10.1017/pasa.2019.42}

\bibitem[{{Wolz} {et~al.}(2014){Wolz}, {Abdalla}, {Blake}, {Shaw}, {Chapman},
  \& {Rawlings}}]{2014MNRAS.441.3271W}
{Wolz}, L., {Abdalla}, F.~B., {Blake}, C., {et~al.} 2014, \mnras, 441, 3271,
  \dodoi{10.1093/mnras/stu792}

\bibitem[{{Zaldarriaga} {et~al.}(2004){Zaldarriaga}, {Furlanetto}, \&
  {Hernquist}}]{2004ApJ...608..622Z}
{Zaldarriaga}, M., {Furlanetto}, S.~R., \& {Hernquist}, L. 2004, \apj, 608,
  622, \dodoi{10.1086/386327}

\bibitem[{{Zhao} {et~al.}(2023){Zhao}, {Mao}, {Zuo}, \&
  {Wandelt}}]{2023arXiv231017602Z}
{Zhao}, X., {Mao}, Y., {Zuo}, S., \& {Wandelt}, B.~D. 2023, arXiv e-prints,
  arXiv:2310.17602, \dodoi{10.48550/arXiv.2310.17602}

\bibitem[{{Zuo} {et~al.}(2019){Zuo}, {Chen}, {Ansari}, \&
  {Lu}}]{2019AJ....157....4Z}
{Zuo}, S., {Chen}, X., {Ansari}, R., \& {Lu}, Y. 2019, \aj, 157, 4,
  \dodoi{10.3847/1538-3881/aaef3b}

\bibitem[{{Zuo} {et~al.}(2023){Zuo}, {Chen}, \& {Mao}}]{2023ApJ...945...38Z}
{Zuo}, S., {Chen}, X., \& {Mao}, Y. 2023, \apj, 945, 38,
  \dodoi{10.3847/1538-4357/acb822}

\end{thebibliography}
\bibliographystyle{aasjournal}

\appendix
\section{GP overview}
\label{subsec: gp overview}

\begin{figure*}
    \centering
    \includegraphics[width=\linewidth]{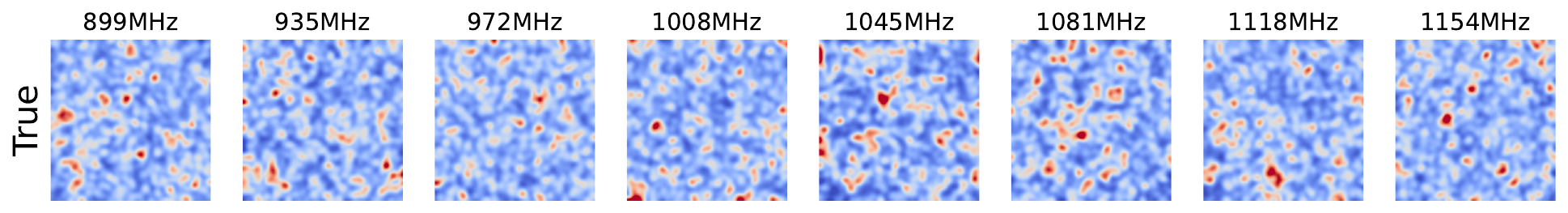}\\
    \includegraphics[width=\linewidth]{CP3Slices.pdf} \\
    \includegraphics[width=\linewidth]{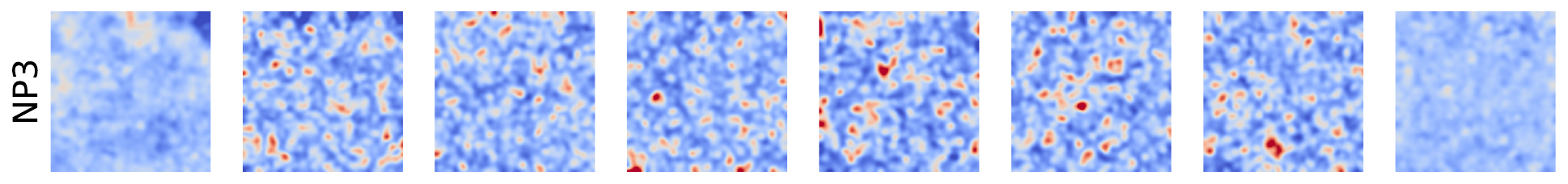} \\
    \includegraphics[width=\linewidth]{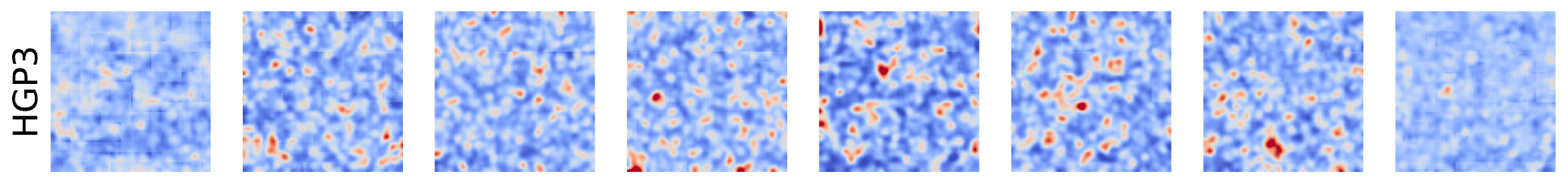} \\
    \caption{{From top to bottom: true {$\bi{f}_{\rm res}$} signal and recovered {$\bi{f}_{\rm res}$} signal by the CP3, NP3 and HGP3 models respectively. From left to right: the {$\bi{f}_{\rm res}$} slices at different frequencies over the full range of the data cube. The color scale is the same as in Figure \ref{fig:visual}. }}
    \label{fig:slices}
\end{figure*}

\begin{figure*}
    \centering
    \includegraphics[width=\linewidth]{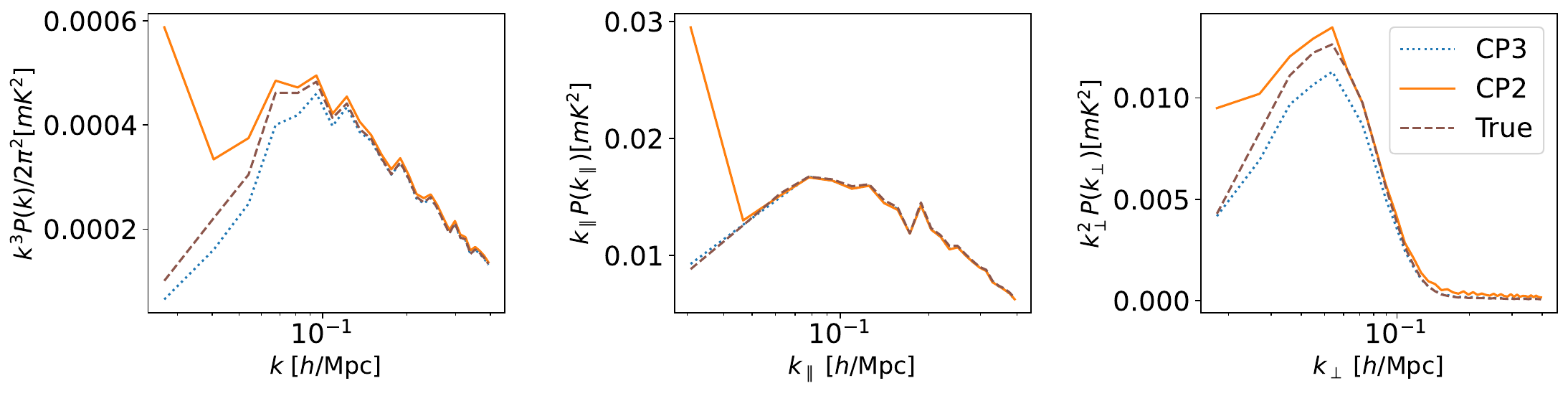}
    \caption{{Same as the upper panel of Figure \ref{fig:psdefault}, but for the CP3 and CP2 models only. The CP2 model overestimates power on large scales due to significant foreground under-cleaning.}}
    \label{fig:PSCP2}
\end{figure*}
GP models are a class of Bayesian non-parametric models that assign a prior distribution over the space of functions \citep{gelman2013bayesian, RasmussenW06}. Broadly, a GP prior consists of an infinite set of random variables for each of the possible inputs of the function. A GP prior may be written as
\begin{equation}
    f\sim\mathcal{GP}(m, {\kappa}),
\end{equation}
where $m$ is the mean function and {$\kappa$ is the GP covariance kernel function}. The mean function defines the function around which realizations from the GP will be distributed, whilst the covariance kernel describes how GP realizations vary around the mean function. The mean function can be decoupled from the GP, writing $f=m+g$ where $g\sim\mathcal{GP}(0, {\kappa})$, and is commonly chosen to be the zero-function. However, informed choice of the mean function can significantly improve regression performance \citep{2012PhRvD..85l3530S, 2022MNRAS.512.1967R}. 

In the setting where we have some finite set of observed data points $y_i,\,i\in\{1,\ldots,N\}$, corresponding to covariates $x_i,\,i\in\{1,\ldots,N\}$, GP regression is made tractable by the fact that the joint distribution over the GP realizations at these points is a multivariate Gaussian with mean $m_i=m(x_i),\,i\in\{1,\ldots,N\}$ and covariance $K_{ij}={\kappa}(x_i,x_j),\,i,j\in\{1,\ldots,N\}$. We may write a generic GP regression model (assuming zero mean function) as,
\begin{equation}
\begin{aligned}
    \Phi&\sim\pi_\Phi(\Phi),\\
    f(x)&\sim\mathcal{GP}(0, K(x,x^\prime|\Phi)),\\
    \sigma_\eta&\sim \pi_{\sigma_\eta}(\sigma_\eta),\\
    y&\sim\pi_y(y|f(x), \sigma_\eta),
    \label{eqn:full model}
\end{aligned}
\end{equation}
where $\Phi$ are the GP covariance parameters with corresponding prior $\pi_\Phi(\Phi)$. The regression function $f(x)$ is assigned a GP prior, $\mathcal{GP}(0, {\kappa}(x,x^\prime|\Phi))$. The parameter $\sigma_\eta$ is chosen to describe the observation noise process and is assigned some prior $\pi_{\sigma_\eta}(\sigma_\eta)$. The data is distributed as $y\sim\pi_y(y|f(x), \sigma_\eta)$, depending on the GP function realizations and the observation noise process. In the case where the likelihood is Gaussian, the GP prior is conjugate to the likelihood, and the GP function realizations can be analytically marginalized. Taking $\sigma_\eta$ to be the Gaussian noise standard deviation, the analytic marginalization yields the simpler model,
\begin{equation}
\begin{aligned}
    \Phi&\sim\pi_\Phi(\Phi),\\
    \sigma_\eta&\sim \pi_{\sigma_\eta}(\sigma_\eta),\\
    y&\sim\mathcal{N}(y|0, K(x,x|\Phi)+\sigma_\eta^2 I_{N}).
    \label{eqn:marginal model}
\end{aligned}
\end{equation}
This amounts to a significant dimensionality reduction, with posterior inference only needing to be performed over the GP covariance and noise parameters. 

For fixed $\Phi$ and $\sigma_\eta$, the expectation and covariance of the GP function realizations at some points $x_*$ may be obtained through the standard prediction formulae,
\begin{equation}
\begin{aligned}
    \mathbb{E}\left[f(x_{*})|\Phi,\sigma_\eta\right]&=K(x_*,x|\Phi)\left(K(x,x|\Phi)+\sigma_\eta^2I_N\right)^{-1}y,\\
    \mathrm{cov}\left[f(x_*)|\Phi,\sigma_\eta\right]&=K(x_*, x_*|\Phi)\\ -K(x_*,x|\Phi)&\left(K(x,x|\Phi)+\sigma_\eta^2I_N\right)^{-1}K(x,x_*|\Phi).
    \label{eqn:prediction formulae}
\end{aligned}
\end{equation}
Posterior samples for $\Phi$ and $\sigma_\eta$ obtained from the marginalized model in Equation~(\ref{eqn:marginal model}) can be used to construct an ensemble estimate for the posterior predictive distribution over $f(x_{*})$ \citep{lalchand2020approximate}. The expected signal is obtained by calculating the ensemble average,
\begin{equation}
    \mathbb{E}\left[f(x_{*})\right] = \frac{1}{M}\sum_{j=1}^{M}\mathbb{E}\left[f(x_{*})|\Phi^j,\sigma_\eta^j\right],
\end{equation}
where the sum is over the $M$ posterior samples, indexed by $j$. The covariance of the posterior predictive samples can be evaluated by drawing samples from the posterior predictive corresponding to each of the posterior samples,
\begin{equation}
    f_{pp}^j(x_{*})\sim\mathcal{N}(\mathbb{E}\left[f(x_{*})|\Phi^j,\sigma_\eta^j\right], \mathrm{cov}\left[f(x_*)|\Phi^j,\sigma_\eta^j\right]),
\end{equation}
and evaluating the ensemble covariance,
\begin{equation}
\begin{aligned}
\mathrm{cov}\left[f(x_{*})\right]=\frac{1}{M-1}\sum_{j=1}^{M}\left\{\left(f_{pp}^j(x_{*}) - \mathbb{E}[f(x_{*})]\right)\right.\\ \left.\otimes \left(f_{pp}^j(x_{*}) - \mathbb{E}[f(x_{*})]\right)\right\}.
\end{aligned}
\end{equation}
Calculating estimators for the GP function realizations in this way ensures that we fully propagate uncertainties in the GP hyperparameters.

\section{Visual Inspection of Different slices}
\label{subsec:vis_insp}

{Figure \ref{fig:slices} displays the true {$\bi{f}_{\rm res}$} signal alongside the reconstructed {$\bi{f}_{\rm res}$} signal obtained by the CP3, NP3 and HGP3 models across different frequency slices. Recovery is worse at lower frequencies, where the foreground signal is brighter. Visual recovery artifacts are particularly apparent for the CP3 and NP3 models at these lower frequencies ($\nu\leq 1008\,\mathrm{MHz}$), most clearly in the top part of the field. Note that all models have poor signal recovery at the very edge of the data cube ($\nu=899\,\mathrm{MHz}$ and $\nu=1154\,\mathrm{MHz}$). This is due to mean reversion of the GP realizations, which is a generic property of GP models. However, this does not have a significant impact on the overall recovery results because these points account for a negligible fraction of the total data cube. Moreover, the uncertainty on the inferred GP realizations at the edges is approximately double that for realizations away from the edges, further down-weighting their impact on the overall results. Improved visual recovery at the edges would likely require physical modeling of the GP mean functions.}

\section{Performance of CP2 model}\label{sec: cp2 performance}

{At the field level the CP2 model was found to have very poor recovery of the {$\bi{f}_{\rm res}$} signal, with typical residuals being $\sim 70\%$ higher compared to the CP3 model. In Figure \ref{fig:PSCP2} we show the power spectrum recovery for the CP2 model. In the middle panel we observe a significant overestimation of power on large scales along the LoS, which implies a pronounced under-cleaning of the foregrounds. These results all render the CP2 model inadequate for \HI{} signal recovery, and hence it was excluded from our more detailed analyses. It is clear that failing to account for polarization leakage in the GP kernel model results in a significant performance degradation.}

\end{document}